# Nonautonomous Volterra Series Expansion of the Variable Phase Approximation and its Application to the Nucleon-Nucleon Inverse Scattering Problem

Gábor Balassa *

*Institute for Particle and Nuclear Physics, HUN-REN Wigner Research Centre for Physics, Konkoly-Thege Miklós út 29-33, 1121 Budapest, Hungary*
*Email: balassa.gabor@wigner.hun-ren.hu

...............................................................................
In this paper, the nonlinear Volterra series expansion is extended and used to describe certain types of nonautonomous differential equations related to the inverse scattering problem in nuclear physics. The nonautonomous Volterra series expansion lets us determine a dynamic, polynomial approximation of the variable phase approximation (VPA), which is used to determine the phase shifts from nuclear potentials through first-order nonlinear differential equations. By using the first-order Volterra expansion, a robust approximation is formulated to the inverse scattering problem for weak potentials and/or high energies. The method is then extended with the help of radial basis function neural networks by applying a nonlinear transformation on the measured phase shifts to be able to model the scattering system with a linear approximation given by the first-order Volterra expansion. The method is applied to describe the $^1S_0$ NN potentials in neutron+proton scattering below 200 MeV laboratory kinetic energies, giving physically sensible potentials and below 1% averaged relative error between the recalculated and the measured phase shifts.
...............................................................................
Subject Index   A13, A34, D00, D06

## 1. Introduction

Many real-life phenomena can be explained by linear or nonlinear dynamical models [1–3], and while linear approximations are adequate to describe such systems in a number of cases, these approximations are usually only able to grasp some of the properties of the full underlying physical system. Some examples where the linear approximation is sufficient can be found in the modeling of thermal systems, in electronics, in structural engineering, or in the linear control of complex systems for small perturbations, etc. On the other hand, many phenomena, e.g. in biology, fluid dynamics, low-temperature thermal systems, etc., can only be described by introducing static or dynamic nonlinearities into the model [4–7], which in turn greatly complicates the analytical and numerical calculations of such problems.

One method to model weak nonlinearities is called the Volterra series approximation [8–10], which is an infinite-dimensional convolutional description of nonlinear systems containing static or dynamic nonlinearities. In comparison to linear models, where the system can be described by a simple convolution of a kernel function and the excitation, in the Volterra representation, the system is described by higher-order convolutional operators in addition to the first-order case. The theory has found its applications in many biological, electrical, and mechanical engineering problems as well [11–16]. Due to its well-developed mathematical foundations, the Volterra approximation can also be used to model the underlying dynamical equations by determining the Volterra kernels straight from the dynamical equations without relying purely on measurements if a mathematical model is at hand.

This paper intends to deal with the problem of inverse scattering in particle and nuclear physics by using the Volterra series approximation on a first-order nonlinear, nonautonomous differential equation called the variable phase approximation (VPA) [17], which is able to describe the asymptotic phase shifts generated by local or nonlocal potentials. Here, we will only deal with local and spherically symmetric potentials; however, the method is much more general and is not limited to only such systems. In general, inverse problems are very hard to solve due to their sensitivity and ill-conditioned nature [18]. In the nuclear inverse scattering problem, a local or nonlocal potential is sought from the asymptotic phase shifts at different energies and/or angular momenta. There are several methods that aim to solve the inverse scattering problem at fixed energies or fixed angular momenta [19,20]; however, a perfect identification of the potential without any preliminary knowledge of the scattering system is almost impossible, and each method gives slightly different results, which also depend on the parameters used in the inversion. Due to this, there is still an ongoing desire to be able to describe the inverse scattering problem with a robust and fast method that is insensitive to at least the measurement noises in the phase shifts.

In this paper, the Volterra series expansion is used to describe the forward problem of elastic two-body quantum scattering given by the VPA. Due to the nonautonomous nature of the forward problem, the original Volterra method has to be extended to be able to describe nonautonomous nonlinear systems as well, which is described in Sect. 2, where first the general model is explained, then a simple example is given where the nonautonomous Volterra kernels are determined analytically. After the model description, the method is put to real use in Sect. 3, where first in Sect. 3.1 the Volterra expansion of the VPA at zero angular momentum (s-wave scattering) is given, then in Sect. 3.2 the convergence of the first-order approximation is studied, while in Sect. 3.3 we study the inversion capabilities of the first-order Volterra expansion with two different basis functions that are used to describe the interaction potentials. After showing the capabilities of the first-order expansion in the inversion procedure, in Sect. 4 the method is applied to real-life s-wave neutron-proton scattering, where due to the more severe nonlinearity, a nonlinear transformation using radial basis function (RBF) neural networks [21] is applied to the phase shifts so that the obtained system could be described as a first-order Volterra model and the inversion equations shown in the previous section could still be applied. At the end, Sect. 5 concludes the paper by mentioning some possible future applications in nuclear and particle physics where the Volterra series could be applied.

## 2. Volterra series approximation for nonautonomous nonlinear systems

In this section, we introduce the Volterra series expansion of autonomous nonlinear dynamical systems and show that, with some modifications, the original Volterra series can be extended to describe nonautonomous nonlinear dynamical systems as well. By using the harmonic balance method, we show a possible way to determine the nonautonomous Volterra kernels in an analytical fashion. The method will then be used to describe a simple nonautonomous nonlinear differential equation by determining the $n$'th-order Volterra kernels in the Laplace domain.

## 2.1. *General model*

Volterra series are originally defined as being able to describe autonomous dynamical systems with weak nonlinearities and have a great many possible applications, mostly in engineering; however, due to their wide modeling capabilities, it is also possible to extend their usage into the fields of nuclear and particle physics. In the original Volterra representation, the output x($r$) of a dynamical system that is excited by an f($r$) function can be described by an infinite-dimensional higher-order convolution integral as:

$$\mathrm{x}(r) \simeq h_0 + \sum_{i=1}^{n} \left[ \underbrace{\int_a^b \ldots \int_a^b}_{i} \mathrm{h}_i(z_1, \ldots, z_i) \prod_{j=1}^{i} \mathrm{f}(r - z_j) dz_j \right], \quad (1)$$

where $r$ is the variable on which the output and input depend, e.g. time, coordinates, etc., $z_i$'s are dummy variables used in the integrals, and $\mathrm{h}_i(z_1, z_2, \ldots, z_i)$ are the $i$'th-order Volterra kernels. The integral limits $a$ and $b$ in general go from $-\infty$ to $\infty$; however, in practice, many real-life physical systems are causal, which means we can take $a = 0$, so in the convolutional integrals we only take into account the "previous" excitations. The upper limit is usually called the memory of the system, which also does not necessarily have to go all to infinity and could have a finite value that corresponds to the memory of the system under description.

The $\mathrm{h}_1(z)$, $\mathrm{h}_2(z_1, z_2)$, ..., $\mathrm{h}_n(z_1, \ldots, z_n)$ kernel functions contain all the information about the dynamics of the system and should have good properties, e.g. continuity, finiteness, boundedness. The nonlinearity comes from the multiplicative terms of the excitation functions, where the number of terms depends on the order of the corresponding kernel. With each added term, a higher nonlinearity could be approximated, and in general, if the corresponding differential equations are known, it should be possible to give general expressions to the analytical form of the Volterra kernels [22,23]. In practice, however, one does not want to keep all the possible higher-order terms, and in many cases, it is sufficient to only keep the second- or third-order terms, which could still give very precise results in practical applications.

In the following, let us assume that $h_0 = 0$, which term will be unimportant in the applications in this paper. The Volterra series representation could also be explained in the Laplace domain as:

$$\tilde{\mathrm{x}}(s) = \mathcal{L}\{\mathrm{x}(r)\}(s) \simeq \tilde{\mathrm{h}}_1(s)\tilde{\mathrm{f}}(s) + \tilde{\mathrm{h}}_2(s_1, s_2)\tilde{\mathrm{f}}(s_1)\tilde{\mathrm{f}}(s_2) + \ldots$$

$$= \sum_{i=1}^{n} \tilde{\mathrm{h}}_i(s_1, \ldots, s_i) \prod_{j=1}^{i} \tilde{\mathrm{f}}(s_j), \quad (2)$$

where $\mathcal{L}\{\cdot\}$ represents the usual $n$-dimensional Laplace transform operator, $\tilde{\mathrm{f}}(s)$ is the Laplace transform of the excitation, and $\tilde{\mathrm{h}}_i(s_1, \ldots, s_i)$ are the Volterra kernels in the Laplace domain. As the convolution operator in the original domain will be a simple product in the Laplace domain, the integral equations are simplified to algebraic equations, which are in many cases more convenient to work with.

One of the easiest ways to obtain the kernel functions straight from the differential equations is through the so-called harmonic balance method [24], where we use specific harmonic functions to excite the system, and by substituting the response of the Volterra representation and the excitations back into the original differential equations, we could express the kernel functions directly by algebraic equations [25]. In this way, we are able to obtain the higher-order kernel functions as well, iteratively, step-by-step, from the kernel functions of the previous order. In the following, the method of obtaining the kernels will be described in detail for autonomous and nonautonomous systems as well.

Firstly, let us assume that we have a general autonomous system, which can be described by a differential equation containing the output x($r$) and its derivatives with respect to the variable $r$, as well as the excitation f($r$). In general, the system can be described in the following form:

$$F\left[\frac{d^k x(r)}{dr^k}, \ldots, \frac{d^2 x(r)}{dr^2}, \frac{dx(r)}{dr}, x(r)\right] = f(r), \quad (3)$$

where on the left-hand side F[·] is a function of the output and its derivatives, while on the right-hand side sits the excitation that drives the system. In general, the F[·] function can be nonlinear as well, which is why we will have to use the Volterra series representation to be able to estimate the solution of the differential equation. The autonomous nature of the system here means that there is no explicit dependence on the variable $r$, which greatly simplifies the determination of the kernel functions.

In the harmonic balance method, let us first use the harmonic excitation $f_1(r) = e^{sr}$, where $s$ is a complex parameter representing the "frequency" of the excitation. Substituting $f_1(r)$ into the Volterra series in Eq. (1) we will arrive at the following infinite sum for the output $x_1(r)$:

$$x_1(r) = \tilde{h}_1(s)e^{sr} + \tilde{h}_2(s,s)e^{2sr} + \ldots + \tilde{h}_n(\underbrace{s, \ldots, s}_{n})e^{nsr}, \quad (4)$$

where the subscript in $x_1(r)$ represents that it is the response for the input $f_1(r)$. The $\tilde{h}_1(s)$, $\tilde{h}_2(s,s)$, ... coefficients are the corresponding transfer functions in the Laplace domain. Let us now substitute $x_1(r)$ back into Eq. (3), where we have to take the derivatives of $x_1(r)$ with respect to $r$ as well. Due to the simple exponential dependence, the $k$'th derivative can be easily expressed as:

$$\frac{d^k x_1(r)}{dr^k} = s^k \tilde{h}_1(s)e^{sr} + 2^k s^k \tilde{h}_2(s,s)e^{2sr} + \ldots + n^k s^k \tilde{h}_n(\underbrace{s, \ldots, s}_{n})e^{nsr}. \quad (5)$$

Substituting everything back, we will arrive at the following algebraic equation:

$$F_1\left[\tilde{h}_1(s), \tilde{h}_2(s,s), \ldots, \tilde{h}_n(s, \ldots, s), e^{sr}, e^{2sr}, \ldots, e^{nsr}, s\right] = e^{sr}, \quad (6)$$

where $F_1[\cdot]$ represents the new function after substituting back the Volterra output for the first harmonic excitation. After collecting the terms with different exponential factors (harmonics), we will arrive at the following structure:

$$u_1(\tilde{h}_1(s), s)e^{sr} + u_2(\tilde{h}_2(s,s), \tilde{h}_1(s), s)e^{2sr} + \ldots + u_n(\tilde{h}_n(s, \ldots, s), \ldots, \tilde{h}_1(s), s)e^{nsr} = 0, \quad (7)$$

where each $u_i(\cdot)$ will be a function of the Volterra kernels and the $s$ variable. To be able to satisfy this equation, each $u_i(\cdot)$ subfunction weighting the different harmonics has to be zero, so $u_i(\tilde{h}_i(s), \ldots, s) = 0$ for all $i$. The general structure will very much depend on the nonlinearity in the original F[·] function; however, by Taylor expansion, it can always be expressed in a polynomial form, so it will always be possible to gather the terms with the corresponding exponential factors even if the system does not have a polynomial nonlinearity. This will be used in later sections, where we need to expand a $\sin^2(\cdot)$ function to be able to determine the Volterra kernels. In general, to determine the kernel functions, we first have to solve an algebraic equation ($u_1(\tilde{h}_1(s), s) = 0$) to obtain the $\tilde{h}_1(s)$ first-order kernel. The other equations do not give us anything useful except the diagonal terms of the higher-order kernels, which is of course not enough as we need the full frequency behavior of the higher-order kernels as well. To do this, let us now excite the system with two harmonics at the same time as $f_2(r) = e^{s_1 r} + e^{s_2 r}$. In this case, the response of the Volterra representation becomes:

$$x_2(r) = \tilde{h}_1(s_1)e^{s_1 r} + \tilde{h}_1(s_2)e^{s_2 r} + 2\tilde{h}_2(s_1, s_2)e^{(s_1+s_2)r}$$
$$+ \tilde{h}_2(s_1, s_1)e^{2s_1 r} + \tilde{h}_2(s_2, s_2)e^{2s_2 r} + \ldots + \text{(higher-order terms)}, \quad (8)$$

where the first two terms correspond to the first-order kernels again, while the third term now corresponds to the second-order kernel at all frequencies $s_1$, and $s_2$ in two dimensions. The other terms are again not important as they do not give us the full frequency behavior. The higher-order terms here consist of the like of $\tilde{h}_3(s_1, s_1, s_2)$ or $\tilde{h}_4(s_1, s_1, s_1, s_1)$ etc., which are also not important. Now putting back $\tilde{x}_2(r)$ into the original differential equation, we will arrive at the following functional representation:

$$F_2\Big[\tilde{h}_1(s_1), \tilde{h}_1(s_2), \tilde{h}_2(s_1, s_1), \tilde{h}_2(s_1, s_2), \tilde{h}_2(s_2, s_2) \ldots, \tilde{h}_n(s_2, \ldots, s_2),$$
$$e^{s_1 r}, e^{2s_1 r}, \ldots, e^{ns_1 r}, e^{s_2 r}, e^{2s_2 r}, \ldots, e^{ns_2 r}, e^{(s_1+s_2)r}, e^{(2s_1+s_2)r},$$
$$e^{(s_1+2s_2)r}, \ldots, e^{(s_1+(n-1)s_2)r}, s_1, s_2\Big] = e^{s_1 r} + e^{s_2 r}. \quad (9)$$

Following the previous arguments, we will obtain algebraic equations for $\tilde{h}_1(s_1)$, $\tilde{h}_1(s_2)$, and $\tilde{h}_2(s_1, s_2, \tilde{h}_1(s_1), \tilde{h}_1(s_2))$, where $\tilde{h}_2(s_1, s_2)$ will also depend on the lower-order kernels, in this case on $\tilde{h}_1(s_1)$ and $\tilde{h}_1(s_2)$. The equations for the first-order kernels will be the same as they were previously, and the only difference now is that they will depend on $s_1$ and $s_2$, and they will appear in the other algebraic equations with these variables. To generalize the procedure to the $n$'th order, we have to excite the system with $f_n(r) = \sum_{i=1}^{n} e^{s_i r}$, in which case the response of the Volterra representation becomes:

$$x_n(r) = \sum_{i=1}^{n} \tilde{h}_1(s_i)e^{s_i r} + \sum_{i_1=1}^{n}\sum_{i_2=i_1}^{n} \mathcal{S}_{i_1 i_2}\tilde{h}_2(s_{i_1}, s_{i_2})e^{(s_{i_1}+s_{i_2})r} + \ldots$$
$$\sum_{i_1=1}^{n}\cdots\sum_{i_n=i_{n-1}}^{n} \mathcal{S}_{i_1\ldots i_n}\tilde{h}_n(s_{i_1}, \ldots, s_{i_n})e^{(\sum_{j=1}^{n} s_{i_j})r}, \quad (10)$$

where $\mathcal{S}_{i_1\ldots i_n}$ are symmetrical factors defined as $n/m!$, where $n$ is the order of the kernel and $m$ is the number of the same frequencies, e.g. for $h_3(s_1, s_1, s_2)$ it is $\mathcal{S}_{112} = 3/2$. Substituting back $x_n(r)$ and $f_n(r)$ into the original differential equation, we get the following functional form:

$$F_n\Big[\tilde{h}_1(s_i), \tilde{h}_2(s_{i_1 i_2}), \ldots, \tilde{h}_n(s_{i_1\ldots i_n}), e^{s_i r}, e^{2s_i r}, \ldots, e^{ns_i r}, e^{(s_{i_1}+s_{i_2})r}, \ldots,$$
$$e^{(s_{i_1}+\ldots+s_{i_n})r}, s_1, s_2, \ldots, s_n\Big] = \sum_{i=1}^{n} e^{s_i r}. \quad (11)$$

After collecting the terms with the same harmonics, we will again have a set of algebraic equations to determine each kernel, starting from $\tilde{h}_1$ up to $\tilde{h}_n$, where each corresponding equation will depend on the results of the previous kernels. To summarize the procedure, to be able to determine the $n$'th-order Volterra kernel, first we have to apply a specific excitation with $n$ different harmonics, and by expressing the response of the Volterra representation to that excitation and substituting it back into the original differential equation, we can express the $n$ algebraic equations by collecting the terms for each harmonic. By solving these equations iteratively, we can obtain each kernel one after another.

The method works well for autonomous systems; however, our main task is to extend the Volterra series representation to nonautonomous systems [22]. In this case, the dynamic equations (the original differential equation) have an explicit dependence on $r$ as well, and the general form can be expressed as:

$$\mathrm{F}\left[\frac{d^k \mathrm{x}(r)}{dr^k}, \ldots, \frac{d^2 \mathrm{x}(r)}{dr^2}, \frac{d\mathrm{x}(r)}{dr}, \mathrm{x}(r), r\right] = \mathrm{f}(r), \qquad (12)$$

where the only difference to Eq. (3) is that now the F[·] function depends on $r$ as well and not just on x($r$) and its derivatives. To be able to generalize this into the Volterra representation, let us assume that the kernel functions in the Laplace domain have an extra parameter dependence (let us call it $R$ for the time being) as follows: $\tilde{\mathrm{h}}_i(s_1, \ldots, s_i) \to \tilde{\mathrm{h}}_i(s_1, \ldots, s_i; R)$. What that means is that now the Volterra kernels in the Laplace domain will be $R$-dependent, therefore, our task is to determine the kernels at all frequencies for all possible $R$ parameters. With the $R$-dependent kernel functions, the response of the Volterra system in the Laplace domain can be expressed as:

$$\tilde{\mathrm{x}}(s; R) \simeq \sum_{i=1}^{n} \tilde{\mathrm{h}}_i(s_1, \ldots, s_i; R) \prod_{j=1}^{i} \tilde{\mathrm{f}}(s_j), \qquad (13)$$

where the $R$ parameter dependence sits in the kernel functions. In practice, we are not interested in just any parameter, but we would like to give a representation where $R = r$ so that the kernel functions in the Laplace domain also depend on the actual $r$ variable. In the following, the meaning of such kernel functions will be clarified, and after the general description, a simple example will be given to show that the method works well for nonautonomous systems.

The first main difference compared to the autonomous case will be in the derivative terms, as now the output of the Volterra system depends on the parameter $R$, which in this case will be equal to the variable $r$, so we are interested in the evolution of the transfer function in the Laplace domain through the evolution of the original $r$ variable. Using this assumption, the first-order derivative of the Volterra response $\mathrm{x}_1(r; R)$ to the $\mathrm{f}_1(r) = e^{sr}$ excitation becomes:

$$\left.\frac{d\mathrm{x}_1(r; R)}{dr}\right|_{R=r} = \frac{d}{dr}\left[\tilde{\mathrm{h}}_1(s; r)e^{sr} + \tilde{\mathrm{h}}_2(s, s; r)e^{2sr} + \ldots\right]$$

$$= \frac{d\tilde{\mathrm{h}}_1(s; r)}{dr}e^{sr} + s\tilde{\mathrm{h}}_1(s; r)e^{sr} + \frac{d\tilde{\mathrm{h}}_2(s, s; r)}{dr}e^{2sr} + 2s\tilde{\mathrm{h}}_2(s, s; r)e^{sr} + \ldots, \quad (14)$$

where it is obvious that due to the $r$-dependence of the transfer functions, the derivative will contain the derivatives of the $\tilde{\mathrm{h}}_i(s; r)$ functions as well as extra contributions. By applying the

$n$'th-order excitations $f_n$ ($n > 1$) the derivative term can be expressed in a generalized form as:

$$\frac{d}{dr}\left[\sum_{i=1}^{n}\tilde{h}_1(s_i;r)e^{s_ir} + \sum_{i_1=1}^{n}\sum_{i_2=i_1}^{n}\mathcal{S}_{i_1i_2}\tilde{h}_{i_1i_2}(s_{i_1},s_{i_2};r)e^{(s_{i_1}+s_{i_2})r} + \ldots\right] =$$

$$\sum_{i=1}^{n}\left(\frac{d\tilde{h}_1(s_i;r)}{dr}e^{s_ir} + s_i\tilde{h}_1(s_i;r)e^{s_ir}\right) +$$

$$\sum_{i_1=1}^{n}\sum_{i_2=i_1}^{n}\mathcal{S}_{i_1i_2}\left(\frac{d\tilde{h}_2(s_{i_1},s_{i_2};r)}{dr}e^{(s_{i_1}+s_{i_2})r} + (s_{i_1}+s_{i_2})\tilde{h}_2(s_{i_1},s_{i_2};r)e^{(s_{i_1}+s_{i_2})r}\right) + \ldots$$

(higher-order terms), (15)

where the higher-order terms again contain the derivatives of the higher-order kernels. The higher-order derivatives are straightforward to express; however, they will not be needed in the applications shown in this paper.

Using the harmonic balance method with the Volterra representation for $x_n$ to the $n$'th-order input $f_n$, we will get a functional representation that depends on the $s$ frequencies and their exponentials $e^{s_1r}, e^{s_2r}, \ldots, e^{s_nr}$ and on the transfer functions $\tilde{h}_1(s_1;r), \tilde{h}_1(s_2;r), \ldots, \tilde{h}_1(s_n;r)$, $\tilde{h}_2(s_1,s_1;r), \tilde{h}_2(s_1,s_2;r), \ldots, \tilde{h}_2(s_n,s_n;r), \ldots, \tilde{h}_n(s_1,s_2,\ldots,s_n;r), \ldots, \tilde{h}_n(s_n,s_n,\ldots,s_n;r)$ and their derivatives with respect to $r$ as follows:

$$G_n\left[\frac{d\tilde{h}_1(s_{i_1};r)}{dr}, \frac{d\tilde{h}_2(s_{i_1},s_{i_2};r)}{dr}, \ldots, \frac{d\tilde{h}_n(s_{i_1},s_{i_2},\ldots,s_{i_n};r)}{dr}, \tilde{h}_1(s_{i_1};r), \tilde{h}_2(s_{i_1},s_{i_2};r), \ldots, \right.$$

$$\left.\tilde{h}_n(s_{i_1},s_{i_2},\ldots,s_{i_n};r), e^{s_1r}, e^{s_2r}, \ldots, e^{s_nr}, s_1, s_2, \ldots, s_n\right] = 0, \quad (16)$$

where the $i_j$ indices in $s_{i_j}$ go from 1 to $n$, with $i_{j+1} \geq i_j$. By expanding $G_n$ we will arrive at a polynomial approximation of the nonlinearities; therefore, it is always possible to collect the exponential terms due to the structure given by the derivatives in Eq. (16). By doing that, we will obtain the following form:

$$\sum_{i=1}^{n}e^{s_ir}\underbrace{g_i(\ldots)}_{=0} + \sum_{i_1=1}^{n}\sum_{i_2=i_1}^{n}e^{(s_{i_1}+s_{i_2})r}\underbrace{g_{i_1i_2}(\ldots)}_{=0} + \ldots + \sum_{i_1=1}^{n}\cdots\sum_{i_n=i_{n-1}}^{n}e^{(\sum_{j=1}^{n}s_{i_j})r}\underbrace{g_{i_1\ldots i_n}(\ldots)}_{=0} = 0, \quad (17)$$

where the $g_i(\ldots)$ subfunctions will be differential equations (possibly nonautonomous), instead of algebraic equations as was the case before, now for the $h_i(\ldots)$ kernels, whose structure will depend on the actual nonlinear differential equations. As before, we have to solve each subequation (now differential equations) to zero to be able to determine the kernel functions iteratively. In the next subsection, we will give explicit expressions for the $g_i(\ldots)$ functions for a simple nonlinear differential equation. After we obtained the kernel functions in the Laplace domain, we could apply the $n$-dimensional inverse Laplace transform to $\tilde{h}_n(s_1, \ldots, s_n; r)$ as:

$$\mathcal{L}_{s_1,\ldots,s_n}\left\{\tilde{h}_n(s_1,\ldots,s_n;r)\right\}(z_1,\ldots,z_n;r) = h_n(z_1,\ldots,z_n;r). \quad (18)$$

Assuming a causal system with $f(r) = 0$ for $r < 0$, the $n$'th-order Volterra response can be expressed as:

$$x(r) \simeq \sum_{i=1}^{n}\left[\underbrace{\int_0^r\cdots\int_0^r}_{i} h_i(z_1,\ldots,z_i;r)\prod_{j=1}^{i}f(r-z_j)dz_j\right], \quad (19)$$

where the convolutional integrals are going from 0 to $r$, and the upper bound $r$ is exactly the extra parameter we introduced into the Laplace-transformed Volterra kernels before. In the next subsection, we will show a simple example of how to deal with a system that is described by a nonautonomous differential equation by calculating the Volterra kernels up to the third order.

### 2.2. *Volterra expansion of a simple nonautonomous nonlinear differential equation*

In this section, we show a simple example of how to determine the nonautonomous Volterra kernels for a nonlinear dynamical system described by the following differential equation:

$$\frac{d\mathrm{x}(r)}{dr} = \mathrm{f}(r)\big(r + \mathrm{x}(r)\big), \tag{20}$$

where $\mathrm{f}(r)$ is the excitation, while $\mathrm{x}(r)$ is the output of the system. Following the description given in the previous subsection, first let us apply a single harmonic excitation $f_1 = e^{sr}$ and determine the corresponding Volterra output, which will be $x_1 = \tilde{\mathrm{h}}_1(s; r)e^{sr} + \tilde{\mathrm{h}}_2(s, s; r)e^{2sr} + \ldots + \tilde{\mathrm{h}}_n(s, \ldots, s; r)$. By substituting this back into Eq. (20) we will arrive at the following differential equation:

$$\frac{d\tilde{\mathrm{h}}_1(s; r)}{dr}e^{sr} + s\tilde{\mathrm{h}}_1(s, r)e^{sr} + \ldots = e^{sr}\big(r + \tilde{\mathrm{h}}_1(s, r)e^{sr} + \tilde{\mathrm{h}}_2(s, s, r)e^{2sr} + \ldots\big), \tag{21}$$

where it is easy to see that by collecting the terms with different harmonics, the differential equation, which corresponds to the $e^{sr}$ terms, will be:

$$\frac{d\tilde{\mathrm{h}}_1(s, r)}{dr} + s\tilde{\mathrm{h}}_1(s, r) = r, \tag{22}$$

which is a first-order linear, nonautonomous differential equation and has the following general solution:

$$\tilde{\mathrm{h}}_1(s; r) = e^{-\int s dr}\left[\int e^{\int s dr} r dr + C_1\right] = \frac{sr - 1}{s^2} + C_1 e^{-sr}, \tag{23}$$

where the $C_1$ constant has to be determined by the initial conditions of the system. Let us assume here, for simplicity, that $\mathrm{x}(r = 0) = 0$, in which case we could choose the integration constant to be $C_1 = 0$, which will be apparent from the inverse Laplace–transformed response. Taking this into account, the first-order Volterra kernel in the Laplace domain can be expressed as:

$$\tilde{\mathrm{h}}_1(s; r) = \frac{sr - 1}{s^2}, \tag{24}$$

which function has a good, continuous, decaying behavior in the $s$-domain as we have expected. The implicit $r$ dependence also shows itself, which tells us that the frequency behavior of the Volterra kernel will change depending on what $r$ we are interested in. To determine the second-order Volterra kernel, let us apply an excitation with two harmonics $s_1$ and $s_2$ in the usual form: $f_2 = e^{s_1 r} + e^{s_2 r}$, giving the Volterra response as $x_2 = \tilde{\mathrm{h}}_1(s_1, r)e^{s_1 r} + \tilde{\mathrm{h}}_1(s_2, r)e^{s_2 r} + 2\tilde{\mathrm{h}}_2(s_1, s_2, r)e^{(s_1+s_2)r} + (\ldots)$. Substituting back into the original differential equations, we will arrive at the following form:

$$\frac{d\tilde{\mathrm{h}}_1(s_1, r)}{dr}e^{s_1 r} + s_1\tilde{\mathrm{h}}_1(s_1, r)e^{s_1 r} + \frac{d\tilde{\mathrm{h}}_1(s_2, r)}{dr}e^{s_2 r} + s_2\tilde{\mathrm{h}}_1(s_2, r)e^{s_2 r}$$

$$+ 2\frac{d\tilde{\mathrm{h}}_2(s_1, s_2, r)}{dr}e^{(s_1+s_2)r} + 2(s_1 + s_2)\tilde{\mathrm{h}}_2(s_1, s_2, r)e^{(s_1+s_2)r} + \ldots$$

$$= \big(e^{s_1 r} + e^{s_2 r}\big)\big(r + \tilde{\mathrm{h}}_1(s_1, r)e^{s_1 r} + \tilde{\mathrm{h}}_1(s_2, r)e^{s_2 r} + 2\tilde{\mathrm{h}}_2(s_1, s_2, r)e^{(s_1+s_2)r} + \ldots\big), \tag{25}$$

where now we have more derivative terms as well as more mixed and higher-order terms. By collecting the different terms with the same harmonics, we can express this in a more convenient form as:

$$e^{s_1 r}\underbrace{\left[\frac{d\tilde{h}_1(s_1,r)}{dr}+s_1\tilde{h}_1(s_1,r)-r\right]}_{g_1(\ldots)}+e^{s_2 r}\underbrace{\left[\frac{d\tilde{h}_1(s_2,r)}{dr}+s_2\tilde{h}_1(s_2,r)-r\right]}_{g_2(\ldots)}+$$

$$e^{(s_1+s_2)r}\underbrace{\left[2\frac{d\tilde{h}_2(s_1,s_2,r)}{dr}+2(s_1+s_2)\tilde{h}_2(s_1,s_2,r)-\tilde{h}_1(s_1,r)-\tilde{h}_1(s_2,r)\right]}_{g_{12}(\ldots)}+\ldots=0, \quad (26)$$

where it can be seen that the differential equations corresponding to the $e^{s_1 r}$ and $e^{s_2 r}$ harmonics are the same as they were before for the $f_1 = e^{sr}$ excitation and again correspond to the first-order Volterra kernels, but now with $s_1$ and $s_2$ variables. The relevant equation for the second-order kernel is shown in the second line with the coefficient $e^{(s_1+s_2)r}$, where it can also be seen that it depends on $\tilde{h}_1(s_1; r)$ and $\tilde{h}_1(s_2; r)$, as stated in the general description of the method in the previous section. Each bracketed term now has to be equal to zero, which gives us basically two differential equations to solve, as the first two equations are the same but with different variables. The first differential equation has been solved previously, while the differential equation for the second-order Volterra kernel has the same structure, and its general solution can be expressed as:

$$\begin{aligned}\tilde{h}_2(s_1,s_2;r)&=e^{-(s_1+s_2)r}\left[\int e^{(s_1+s_2)r}\left(\frac{\tilde{h}_1(s_1;r)+\tilde{h}_1(s_2;r)}{2}\right)dr+C_2\right]\\&=e^{-(s_1+s_2)r}\left[\int e^{(s_1+s_2)r}\left(\frac{s_1 r-1}{2s_1^2}+\frac{s_2 r-1}{2s_2^2}\right)dr+C_2\right]\\&=\frac{r(s_1^2 s_2+s_2^2 s_1)-s_1^2-s_2^2-s_1 s_2}{2s_1^2 s_2^2(s_1+s_2)}+C_2 e^{-(s_1+s_2)r},\end{aligned} \quad (27)$$

where on the second line we substituted the solutions for $\tilde{h}_1(s_1; r)$ and $\tilde{h}_1(s_2; r)$. The solution again has an extra constant, $C_2$, which can be set to zero due to the same reasoning as before. The solution for the second-order nonautonomous Volterra kernel can therefore be expressed as:

$$\tilde{h}_2(s_1,s_2;r)=\frac{r(s_1^2 s_2+s_2^2 s_1)-s_1^2-s_2^2-s_1 s_2}{2s_1^2 s_2^2(s_1+s_2)}. \quad (28)$$

Before we give a general expression for the determination of the $n$'th-order kernels, let us work out the third-order kernels by applying the usual excitation but now with three harmonics as: $f_3 = e^{s_1 r} + e^{s_2 r} + e^{s_3 r}$. The response of the Volterra system now will be $x_3 = \tilde{h}_1(s_1,r)e^{s_1 r} + \tilde{h}_1(s_2,r)e^{s_2 r} + \tilde{h}_1(s_3,r)e^{s_3 r} + 2\tilde{h}_2(s_1,s_2,r)e^{(s_1+s_2)r} + 2\tilde{h}_2(s_1,s_3,r)e^{(s_1+s_3)r} + 2\tilde{h}_2(s_2,s_3,r)e^{(s_2+s_3)r} + 6\tilde{h}_3(s_1,s_2,s_3,r)e^{(s_1+s_2+s_3)r} + (\ldots)$. Putting everything back into Eq. (20)

we will arrive at the following form:

$$\frac{d\tilde{h}_1(s_1, r)}{dr}e^{s_1 r} + s_1\tilde{h}_1(s_1, r)e^{s_1 r} + \frac{d\tilde{h}_1(s_2, r)}{dr}e^{s_2 r} + s_2\tilde{h}_1(s_2, r)e^{s_2 r} + \frac{d\tilde{h}_1(s_3, r)}{dr}e^{s_3 r}$$
$$+ s_3\tilde{h}_1(s_3, r)e^{s_3 r} + 2\frac{d\tilde{h}_2(s_1, s_2, r)}{dr}e^{(s_1+s_2)r} + 2(s_1 + s_2)\tilde{h}_2(s_1, s_2, r)e^{(s_1+s_2)r}$$
$$+ 2\frac{d\tilde{h}_2(s_1, s_3, r)}{dr}e^{(s_1+s_3)r} + 2(s_1 + s_3)\tilde{h}_2(s_1, s_3, r)e^{(s_1+s_3)r} + 2\frac{d\tilde{h}_2(s_2, s_3, r)}{dr}e^{(s_2+s_3)r}$$
$$+ 2(s_2 + s_3)\tilde{h}_2(s_2, s_3, r)e^{(s_2+s_3)r} + + 6\frac{d\tilde{h}_3(s_1, s_2, s_3, r)}{dr}e^{(s_1+s_2+s_3)r}$$
$$+ 6(s_1 + s_2 + s_3)\tilde{h}_3(s_1, s_2, s_3, r)e^{(s_1+s_2+s_3)r} + \ldots = \left[e^{s_1 r} + e^{s_2 r} + e^{s_3 r}\right]$$
$$\times \left[r + \tilde{h}_1(s_1, r)e^{s_1 r} + \tilde{h}_1(s_2 r)e^{s_2 r} + \tilde{h}_1(s_3, r)e^{s_3 r} + 2\tilde{h}_2(s_1, s_2, r)e^{(s_1+s_2)r}\right.$$
$$\left. + 2\tilde{h}_2(s_1, s_3, r)e^{(s_1+s_3)r} + 2\tilde{h}_2(s_2, s_3, r)e^{(s_2+s_3)r} + 6\tilde{h}_3(s_1, s_2, s_3, r)e^{(s_1+s_2+s_3)r} + \ldots\right], \quad (29)$$

where now we have terms for the determination of the $\tilde{h}_3(s_1, s_2, s_3)$ kernel functions as well. The expression looks a little tedious; however, by collecting the different terms, we will see that it has a very general form. But before we generalize it to the $n$'th order, let's express the three relevant differential equations for the first-, second-, and third-order kernels. By collecting the terms with the same harmonics, we arrive at the following equation:

$$\sum_{i=1}^{3} e^{s_i r}\left[\frac{d\tilde{h}_1(s_i, r)}{dr} + s_i\tilde{h}_1(s_i, r) - r\right]$$
$$+ \sum_{i=1}^{3}\sum_{j=i}^{3} e^{(s_i+s_j)r}\left[\mathcal{S}_{ij}\frac{d\tilde{h}_2(s_i, s_j, r)}{dr} + \mathcal{S}_{ij}(s_i + s_j)\tilde{h}_2(s_i, s_j, r) - \tilde{h}_1(s_i, r) - \tilde{h}_1(s_j, r)\right]$$
$$+ \sum_{i=1}^{3}\sum_{j=i}^{3}\sum_{k=j}^{3} e^{(s_i+s_j+s_k)r}\left[\mathcal{S}_{ijk}\frac{d\tilde{h}_3(s_i, s_j, s_k, r)}{dr} + \mathcal{S}_{ijk}(s_i + s_j + s_k)\tilde{h}_3(s_i, s_j, s_k, r)\right.$$
$$\left. - \mathcal{S}_{ij}\tilde{h}_2(s_i, s_j, r) - \mathcal{S}_{ik}\tilde{h}_2(s_i, s_k, r) - \mathcal{S}_{jk}\tilde{h}_2(s_j, s_k, r)\right] + \ldots = 0, \quad (30)$$

where again, we will have higher-order but irrelevant terms as well. As before, all the bracketed terms have to be zero, which gives us the differential equations to solve for the kernel functions up to the third order. The differential equations for the first-order kernels will have the exact same form as before:

$$\frac{d\tilde{h}_1(s_i; r)}{dr} + s_i\tilde{h}_1(s_i; r) - r = 0, \quad (31)$$

where $i = 1, 2, 3$. The differential equations for the second-order kernels now will be:

$$\mathcal{S}_{ij}\frac{d\tilde{h}_2(s_i, s_j; r)}{dr} + \mathcal{S}_{ij}(s_i + s_j)\tilde{h}_2(s_i, s_j; r) - \tilde{h}_1(s_i; r) - \tilde{h}_1(s_j; r) = 0, \quad (32)$$

with $i \neq j$ so $\mathcal{S}_{ij} = 2$. To determine the third-order kernels, the only relevant term is when $i \neq j \neq k$, so $(i, j, k) = (1, 2, 3)$, therefore $\mathcal{S}_{ijk} = 6$, and $\mathcal{S}_{ij} = \mathcal{S}_{ik} = \mathcal{S}_{jk} = 2$, in which case we will arrive at the following differential equation:

$$6\frac{d\tilde{h}_3(s_1, s_2, s_3; r)}{dr} + 6(s_1 + s_2 + s_3)\tilde{h}_3(s_1, s_2, s_3; r) =$$
$$2\tilde{h}_2(s_1, s_2; r) + 2\tilde{h}_2(s_1, s_3; r) + 2\tilde{h}_2(s_2, s_3; r). \tag{33}$$

Following the expressions given for the higher-order kernels, it is easy to generalize to the $n$'th order, in which case the differential equations for the nonautonomous Volterra kernels can be expressed in the following compact form:

$$\frac{d\tilde{h}_n(s_1, \ldots, s_n, r)}{dr} + \left(\sum_{i=1}^{n} s_i\right) \tilde{h}_n(s_1, \ldots, s_n, r) = \frac{(n-1)!}{n!} \sum_{i_1, i_2, \ldots, i_{n-1} \in \mathcal{P}_n} \tilde{h}_{n-1}(s_{i_1}, \ldots, s_{i_{n-1}}, r), \tag{34}$$

where the factorial coefficients on the right-hand side are coming from the statistical factors $\mathcal{S}$, and the $i_1, \ldots, i_{n-1} \in \mathcal{P}_n$ in subscript in the sum on the right-hand side means that we take all the possible unique permutations with $i_j < i_k$ for $j < k$ of the variables $s_1, \ldots, s_n$ for the $n-1$'th kernel, e.g. for $n = 4$ the right-hand side sum will consist of the terms $\tilde{h}_3(s_1, s_2, s_3; r)$, $\tilde{h}_3(s_1, s_2, s_4; r)$, $\tilde{h}_3(s_1, s_3, s_4; r)$, and $\tilde{h}_3(s_2, s_3, s_4; r)$.

To be able to give a cleaner representation of the results, let's go back to the "original" domain by inverse Laplace transforming the kernel functions by the definition of Eq. (18), where "$r$" is just a parameter. The first-order kernel then becomes:

$$h_1(z; r) = r - z, \tag{35}$$

where the $r$ dependence is obvious. Applying the 2D inverse Laplace transform to $h_2(s_1, s_2; r)$ we will arrive at the following expression:

$$h_2(z_1, z_2; r) = \frac{r}{2} + (z_1 - z_2)\Theta(z_2 - z_1) - \frac{z_1}{2}, \tag{36}$$

where $\Theta(z_2 - z_1)$ is the Heaviside function. Finally, the inverse Laplace transform of the third-order Volterra kernel becomes:

$$h_3(z_1, z_2, z_3; r) = \left[\Theta(z_2 - z_1)\Theta(z_3 - z_1)\left(\frac{z_3}{6} - \frac{z_1}{6}\right)\right.$$
$$+ \Theta(z_2 - z_1)\Theta(z_3 - z_2)\left(\frac{z_2}{6} - \frac{z_3}{6}\right)$$
$$\left.+ \Theta(z_2 - z_1)\left(\frac{z_1}{6} - \frac{z_2}{6}\right) + \Theta(z_3 - z_1)\left(\frac{z_1}{6} - \frac{z_3}{6}\right) - \frac{z_1}{6}\right] + \frac{r}{6}, \tag{37}$$

where again we have a simple $r$ dependence as well. The response of the system can now be expressed in the following convolutional integral form:

$$x(r) \simeq \int_0^r h_1(z, r)f(r - z)dz + \int_0^r \int_0^r h_2(z_1, z_2, r)f(r - z_1)f(r - z_2)dz_1 dz_2$$
$$+ \int_0^r \int_0^r \int_0^r h_3(z_1, z_2, z_3, r)f(r - z_1)f(r - z_2)f(r - z_3)dz_1 dz_2 z_3 + \ldots, \tag{38}$$

where, due to the simple linear dependence on the $z_i$ dummy variables, it is very easy to calculate the Volterra response for, e.g. polynomials; therefore, it is possible to give analytical expressions for a wide range of excitations.

To show the behavior of the Volterra approximations of different orders, let us force the system with a constant excitation: $f(r) = A$. To proceed further, let us assume that the differential equation in Eq. (20) is in a dimensionless form, therefore, $r$, x($r$), and f($r$) are all dimensionless quantities. Let's also set the amplitude to $A = 3$, and calculate the first-, second-, and third-order Volterra approximations, which can be integrated out in closed form, giving the following simple polynomial approximation to the solution of the nonlinear dynamical system:

$$\text{x}(r) \simeq Ar^2 + A^2 r^3 \left(\frac{1}{4} - \frac{1}{6}\right) + A^3 r^4 \left(\frac{1}{12} + \frac{1}{48} - \frac{1}{144} - \frac{1}{18}\right) + \mathcal{O}(r^5), \tag{39}$$

where the different coefficients in the brackets come from the integration of the separate parts with the Volterra kernels in Eqs. (35), (36), and (37). The results of the approximation can be seen in Fig. 1, where the numerical solution of Eq. (20) is also shown with the first three Volterra approximations. The results show that the Volterra approximation is better with increasing order, which is the exact behavior that we wanted to see. It is also evident that the nonlinearity in the original equation and/or the amplitude of the excitation are too large to be able to describe the system in this range. While modeling a specific system, it is always crucial to check the operating range of the specific approximation, e.g. first-order Volterra, which in this case would mean the amplitude, type, and range of the excitations. In the next section, we will go a step further and apply the nonautonomous Volterra expansion to a specific nonlinear differential equation, which is able to describe the phase shifts in two-body quantum scattering experiments. Therefore, we will obtain an approximation that could be used in real-world scattering problems.

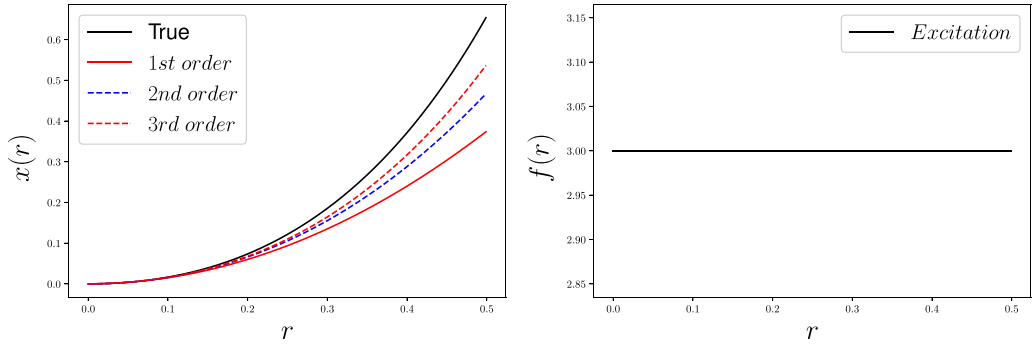

**Fig. 1.** The first three Volterra approximations (left) of the nonautonomous, nonlinear differential equation in Eq. (20) with $A = 3$ constant excitation (right).

## 3. Volterra expansion of the VPA at zero angular momentum

In this section, we will apply the nonautonomous Volterra expansion to the so-called VPA, which is a first-order, nonautonomous, nonlinear differential equation and describes the phase shift evolution through the interaction potential in two-body elastic scattering. In general, to be able to describe the phase shifts in scattering experiments, one has to solve the Schrodinger equation for an incoming plane wave in coordinate space until it reaches an asymptotic value. In contrast, the VPA equation describes the evolution of the phase function through the particle's trajectory through the interaction potential, while the asymptotic value of the phase function corresponds to the measurable phase shifts. The VPA equation for general angular momentum can be expressed as [17]:

$$\frac{d\delta_l(r,k)}{dr} = -\frac{2\mu V(r)}{k\hbar^2} \Big[ \text{j}_l(kr) \cos(\delta_l(r,k)) - \text{n}_l(kr) \sin(\delta_l(r,k)) \Big]^2, \tag{40}$$

where $V(r)$ is the interaction potential, $\delta_l(r)$ is the phase function at distance $r$, and at momentum $k$ defined as $k = (2\mu E/\hbar^2)^{1/2}$, while $j_l(kr)$ and $n_l(kr)$ are the Ricatti–Bessel and Ricatti–Neumann functions, respectively.

The VPA is just a reformulation of the same quantum mechanical two-body problem which can be described by the Schrodinger equation, and while it may have been more sensible to look at the scattering problem through the evolution of the phase shift, it is still not an easy task to solve analytically or numerically, especially for higher angular momenta, where usually 4th- or 5th-order Runge–Kutta methods are necessary [25]. In the following subsections, we will use the previously described nonautonomous Volterra method to approximate the solution of the VPA equation for $l = 0$, which describes the elastic s-wave scattering of two particles. In particular, we will be interested in the so-called best linear approximation (BLA) [26], which is described by the first-order Volterra system. After the determination of the Volterra kernels, we will apply a simple convergence study to be able to determine the operating range of the BLA, which is necessary to use the method for our main task, which is the construction of a robust inversion method.

### 3.1. *Determination of the Volterra kernels*

From now on, we will be interested only in s-wave scattering, therefore, in Eq. (40) we set the angular momentum to $l = 0$, which greatly simplifies the equation to the following form:

$$\frac{d\delta_0(r,k)}{dr} = -\frac{2\mu V(r)}{k\hbar^2} \sin^2(kr + \delta_0(r,k)), \tag{41}$$

where again $V(r)$ is the interaction potential, $\mu$ is the reduced mass of the two-body systems, $k$ is the center-of-mass momentum, and $\delta_0(r,k)$ is the $l = 0$ phase shift. This differential equation still has a dominant $\sin^2(\cdot)$ nonlinearity with the variable $r$ sitting inside, which makes the system essentially nonautonomous, which greatly complicates the analytical solution of such a problem. Previously, we assumed that the nonlinearity in the dynamical equations was polynomial, therefore, we can express the Volterra system easily using the harmonic balance method by collecting the different harmonics and solving the resulting differential equations for the Volterra kernels. To be able to use the same method to determine the kernels, first we will Taylor expand the nonlinear term as:

$$\sin^2(x) \simeq x^2 - \frac{x^4}{3} + \frac{2x^6}{45} - \frac{x^8}{315} + \mathcal{O}(x^9) = \sum_{i=1}^{\infty} \frac{(-1)^{i+1} 2^{2i-1} x^{2i}}{(2i)!}, \tag{42}$$

where $x = kr + \delta_0(r)$; therefore, the nonlinearity can be expressed further as:

$$\sin^2(kr + \delta_0(r,k)) \simeq C_1 (kr + \delta_0(r,k))^2 + C_2 (kr + \delta_0(r,k))^4 + \ldots =$$

$$= \sum_{i=1}^{\infty} C_i (kr + \delta_0(r,k))^{2i}, \tag{43}$$

where for simplicity, we have defined the $C_i$ coefficients as:

$$C_i = \frac{(-1)^{i+1} 2^{2i-1}}{(2i)!}. \tag{44}$$

The series expansion of the nonlinear term introduces another approximation into the model, and it is necessary to check the number of dominant terms we need to be able to describe the system in a desired operating range. This will be done in the next subsection, where the convergence analysis of the BLA will be carried out. It is also worth mentioning that instead of Taylor expanding the nonlinearity, it might have been better to express it by some orthogonal polynomial, e.g. Chebyshev or Legendre, in which case the convergence could have been better. Putting the Taylor expanded form back into the VPA equation in Eq. (42) we will arrive at the following differential equation:

$$\frac{d\delta_0(r,k)}{dr} \simeq -\frac{2\mu V(r)}{k\hbar^2}\left[\sum_{i=1}^{N} C_i\big(kr + \delta_0(r,k)\big)^{2i}\right], \qquad (45)$$

where we have truncated the Taylor series expansion at order $N$ and therefore approximated the nonlinearity at $\mathcal{O}(x^{2N+1})$.

Let us now proceed and determine the first-order Volterra kernel by applying $V_1 = e^{sr}$ excitation, in which case the response of the Volterra system will be in the usual $\tilde{h}_1(s;r,k)e^{sr} + \tilde{h}_2(s,s;r,k)e^{2sr} + (\ldots)$ form, where the transfer functions will also depend on the $k$ momentum variable. Substituting this expression into Eq. (45), then taking its derivative on the left-hand side, we arrive to the following equation for the transfer functions:

$$\frac{d\tilde{h}_1(s;r,k)}{dr}e^{sr} + s\tilde{h}_1(s;r,k)e^{sr} + \ldots = -\frac{2\mu V_1(r)}{k\hbar^2}\left[\sum_{i=1}^{N} C_i\big(kr + \tilde{h}_1(s;r,k)e^{sr} + \ldots\big)^{2i}\right], \quad (46)$$

where we have omitted the higher-order terms, which have no importance in the determination of the first-order transfer functions. Now let's express the right-hand side by using the binomial theorem as:

$$(kr + \tilde{h}_1(s;r,k)e^{sr} + \ldots)^{2i} = \sum_{j=0}^{2i}\binom{2i}{j}(kr)^j\big(\tilde{h}_1(s;r,k)e^{sr} + \ldots\big)^{2i-j}$$

$$= \sum_{j=0}^{2i}\binom{2i}{j}(kr)^j\big(\tilde{h}_1(s;r,k)e^{sr} + \tilde{h}_2(s,s;r,k)e^{2sr} + \ldots + \tilde{h}_n(s,\ldots,s;r,k)e^{nsr}\big)^{2i-j},$$

(47)

which is a necessary step to be able to separate the different harmonics and collect the relevant terms. In the Taylor expansion, the only contributions to the $e^{sr}$ terms are $C_1(kr)^2 + C_2(kr)^4 + C_2(kr)^6 + \ldots$ due to the multiplicative relation with the potential (excitation $V_1(r)$), which adds an extra $e^{sr}$ term to the mixed and to the $(\delta_0)^n$ terms, which will consist of at least an $e^{sr}$ factor, so even at the lowest order, the mixed terms will give a contribution to the $e^{nsr}$ ($n > 1$) terms, which are not needed in the determination of the first-order kernels. Here, this means only the $j = 2i$ term will contribute to the determination of the first-order kernel. This can be seen more clearly by writing out all the necessary terms in Eq. (46), in which case we will arrive at the following form:

$$\frac{d\tilde{h}_1(s;r,k)}{dr}e^{sr} + s\tilde{h}_1(s;r,k)e^{sr} + \ldots + \frac{d\tilde{h}_n(s,\ldots,s;r,k)}{dr}e^{nsr} + ns\tilde{h}_n(s,\ldots,s;r,k)e^{sr} =$$

$$-\frac{2\mu e^{sr}}{k\hbar^2}\left[\sum_{i=1}^{N} C_i \sum_{j=0}^{2i}\binom{2i}{j}(kr)^j\big(\tilde{h}_1(s;r,k)e^{sr} + \ldots + \tilde{h}_n(s,\ldots,s;r,k)e^{nsr}\big)^{2i-j}\right]. \quad (48)$$

Collecting the first-order terms with the $e^{sr}$ harmonics and equating them to zero gives the following first-order nonautonomous differential equation for the first-order kernels:

$$\frac{d\tilde{h}_1(s;r,k)}{dr} + s\tilde{h}_1(s;r,k) = -\frac{2\mu}{k\hbar^2}\sum_{i=1}^{N} C_i(kr)^{2i}. \qquad (49)$$

The general solution to this differential equation can be given as:

$$\tilde{h}_1(s; r, k) = -e^{-\int s dr} \left\{ \int \left[ e^{\int s dr} \left\{ \frac{2\mu}{\hbar^2} \sum_{i=1}^{N} C_i k^{2i-1} r^{2i} \right\} \right] dr + K \right\}, \quad (50)$$

where the momentum $k$ outside the sum has been put inside the sum, giving an overall $k^{2i-1}$ contribution to the right-hand side of the differential equation. To solve the integral, we can use the following general expression for the integral of an exponential multiplied by a polynomial:

$$\int e^{sr} r^{2i} dr = \frac{(-1)^{2i} \Gamma(2i+1, -rs)}{s^{2i+1}}, \quad (51)$$

where $\Gamma(2i+1, -rs)$ is the incomplete Gamma function, which in the case where $2i+1$ is an integer can be given by the following sum:

$$\Gamma(2i+1, -rs) = e^{rs} (2i)! \sum_{j=0}^{2i} \frac{(-1)^{2i-j}}{(2i-j)!} (rs)^{2i-j}. \quad (52)$$

Substituting back the results into Eq. (50) and omitting the $K$ integration constant due to the initial conditions of the VPA equation, we arrive at the general expression for the first-order nonautonomous Volterra kernel in the Laplace domain:

$$\tilde{h}_1(s; r, k) = \frac{2\mu}{\hbar^2} \sum_{i=1}^{N} \sum_{j=0}^{2i} \left\{ C_i \frac{(-1)^{j+1}(2i)!}{(2i-j)!} \frac{k^{2i-1}}{s^{j+1}} r^{2i-j} \right\}, \quad (53)$$

where we used the fact that $(-1)^{4i-j} = (-1)^j$ and the overal minus sign goes into the sum, giving an overall $(-1)^{j+1}$ sign factor. The next step is to apply the inverse Laplace transform to Eq. (53) to be able to easily express the phase shifts in coordinate space, which is more suitable for the applications in which we would like to use the method.

As the first-order kernels in the Laplace domain have a simple $1/s^{j+1}$ dependence on the frequencies, the inverse Laplace transform can be carried out analytically by using the following transformation rule [27]:

$$\mathcal{L}_z \left\{ \frac{1}{s^{j+1}} \right\} = \frac{z^j}{j!}. \quad (54)$$

Applying the inverse transform, the first-order nonautonomous Volterra kernel can be expressed in the following form:

$$h_1(z; r, k) = \frac{2\mu}{\hbar^2} \sum_{i=1}^{N} \sum_{j=0}^{2i} \left\{ C_i \frac{(-1)^{j+1}(2i)!}{j!(2i-j)!} k^{2i-1} r^{2i-j} z^j \right\}$$

$$= \frac{2\mu}{\hbar^2} \sum_{i=1}^{N} \sum_{j=0}^{2i} \left\{ \frac{(-1)^{i+j} 2^{2i-1}}{j!(2i-j)!} k^{2i-1} r^{2i-j} z^j \right\}, \quad (55)$$

where in the second line we have substituted $C_i$ from Eq. (44) and used the fact that $(-1)^{i+j+2} = (-1)^{i+j}$.

In the next subsection, the applicability of the BLA will be studied through a simple convergence analysis, where we check the number of necessary terms in the Taylor expansion of the nonlinearity and also address the operating range of the first-order Volterra approximation.

### 3.2. *Convergence of the first-order approximation*

Let us first address the Taylor expansion of the $\sin^2()$ nonlinearity in Eq. (41) by making some assumptions on the problem we would like to describe with the Volterra approximation. As the aim of this study is to describe low-energy, elastic nuclear scattering, we could make some safe assumptions, e.g. on the range and strength of the expected potentials [28,29]. Before that, however, let's check the convergence radius of the Taylor expansion, which can be easily done by the ratio test as follows:

$$\lim_{N\to\infty}\left|\frac{\frac{(-1)^{N+2}2^{2N+1}x^{2N+2}}{(2N+2)!}}{\frac{(-1)^{N+1}2^{2N-1}x^{2N}}{(2N)!}}\right| = \lim_{N\to\infty}\left|\frac{4x^2}{(2N+2)(2N+1)}\right| = 0, \tag{56}$$

where $x = kr + \delta(r)$, and the results indicate that the convergence radius of the Taylor expansion of the $\sin^2()$ function is infinite, thus we could safely use it to describe the nonlinearity in the VPA equation. This, however, does not tell us the number of necessary terms we have to use to estimate the original differential equation with good accuracy. To assess this, we will check the error of the $N$'th-order Taylor expansion of $f(x) = \sin^2(x)$ near $x_0 = 0$, which can be characterized by the $N + 1$'th derivative of $f(x)$ as:

$$R_N(x) = \left.\frac{d^{N+1}f(x)}{dx^{N+1}}\right|_{x=\zeta}\frac{(x-x_0)^{N+1}}{(N+1)!} = \sin^2(\zeta)\frac{x^{N+1}}{(N+1)!}, \tag{57}$$

where $\zeta$ is just some value where we would like to express the function, while we set $x_0 = 0$. Due to the bounded nature of the nonlinearity, an upper bound on the absolute value of the error can be given as follows:

$$|R_N(x)| = \left|\sin^2(\zeta)\frac{x^{N+1}}{(N+1)!}\right| \leq \underbrace{\left|\frac{x^{N+1}}{(N+1)!}\right|}_{E_N(x)}, \tag{58}$$

where $E_N(x)$ represents the upper bound of the absolute error in the $N$'th-order Taylor expansion of the nonlinearity. In Fig. 2 the error dependence on $x$ and $N$ can be followed, where $x$ is not just the distance, but the combination of the distance, momentum, and the coordinate-dependent phase shifts as well. As the phase shift can be negative or positive, a rapidly changing phase function could make $x$ smaller (or at least not grow too much with increasing $r$); therefore, that would make the error of the Taylor approximation smaller or at least more controlled as well. In the worst case, if the potential is, e.g. strictly negative, the phase shifts would be a monotonically increasing positive function, therefore, $x$ will be a monotonically increasing function as well. This would mean that we would need more and more terms in the Taylor expansion as we go further in $r$. In nuclear physics, $r$ is a few fermi, and $k$ depends on the bombarding energy, and the masses of the colliding particles. For nucleon–nucleon collisions at a few MeV up to a few hundreds of MeV it is around $k \in [0.1, 1]$ fm$^{-1}$, while the phase shift also depends on the number of bound states, etc. An example for the possible range we are dealing with is $r \in [0, 5]$ fm, $k \in [0.1, 1]$ fm$^{-1}$, while $\phi(r) \approx [-\pi, \pi]$ as we do not want to include many bound states in the calculations. By trying out the Volterra approximation for many bounded, vanishing potentials, it has been concluded that $N = 20$ will be more than enough for practical calculations.

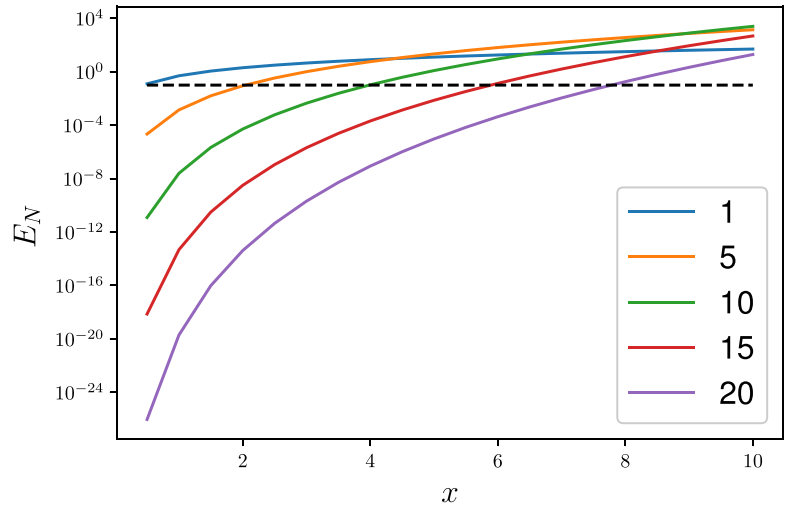

**Fig. 2.** Absolute error of the Taylor expansion of the $\sin^2(x)$ nonlinearity defined in Eq. (58). The colored curves with different colors show the errors at different orders, while the dashed black line at $E_N = 0.1$ is only for reference.

The first-order Volterra approximation is, of course, not the full description of the nonlinear dynamical system, therefore, it is also necessary to check what the operating range of the model is, which in this case means the range of the potentials in both magnitude and distance. This is also strongly related to the previously examined Taylor expansion, therefore, the number of terms in the expansion should also depend on the possible potentials. To get a grasp on the operating range of this system, we will generate many vanishing potentials with a predefined range and magnitude and check the relative errors of the first-order Volterra approximation, defined as:

$$\Delta(A, k) = \frac{1}{N_S} \sum_{i=1}^{N_S} \frac{|\delta_0^{(i)}(\hat{r}, k) - \delta_V^{(i)}(\hat{r}, k)|}{|\delta_0^{(i)}(\hat{r}, k)|}, \tag{59}$$

where $A$ represents the magnitude, $N_S$ is the number of samples, and $\delta_0^{(i)}(r_j)$ is the true phase function of the $i$'th sample at coordinate $\hat{r}$ calculated directly by the VPA equation, while $\delta_V^{(i)}(r_j)$ is the first-order Volterra approximation of the phase function calculated by convolving the potentials with the $h_1(z; r)$ kernel functions in Eq. (55). The relative errors are only calculated at the "asymptotic" value $\hat{r}$ set to be 5 fm according to the test potentials' maximum range. To generate the test potentials, a simple random-phase multisine form is used [30], which can be defined as:

$$V(r; A, L) = A e^{-r/L} \frac{\sum_{n=1}^{N} \sin(n\omega_0 r + \phi_n)}{\mathcal{N}}, \tag{60}$$

where $\mathcal{N}$ is a normalization factor and is set to be $\mathcal{N} = \max(|\sum_{n=1}^{N} \sin(n\omega_0 r + \phi_n)|)$, while $A$ is the "amplitude" of the potential, $N$ is chosen randomly for each sample from $N \in U[\ldots]$, and $\phi_n$ is sampled uniformly from the interval $\phi_n \in U[0, \pi]$, where $U$ represents the uniform distribution. The exponential factor makes sure that the multisine signal vanishes after a certain distance set by $L$. The parameters applied here generate continuous, "not too oscillating," and vanishing excitations in the range $V(r) \in [-A, A]$ MeV. In Fig. 3 the averaged relative errors for the multisine test potentials at three different laboratory energies are shown by using $N_S = 1000$ samples in the error estimation. From the relative errors shown in Fig. 3 we can crudely estimate the operating range of the first-order Volterra approximation; however, one has to keep in mind that these errors are calculated by using only vanishing multisine test potentials in the range of $r \approx [0, 5]$ [fm], which is sufficient to describe the physical potentials we seek in scattering experiments, but it might not be valid for other systems. By using these assumptions, we could deduce that the first-order Volterra approximation could describe the scattering system with a few percent accuracy when the potentials are below a few tens of MeV and the

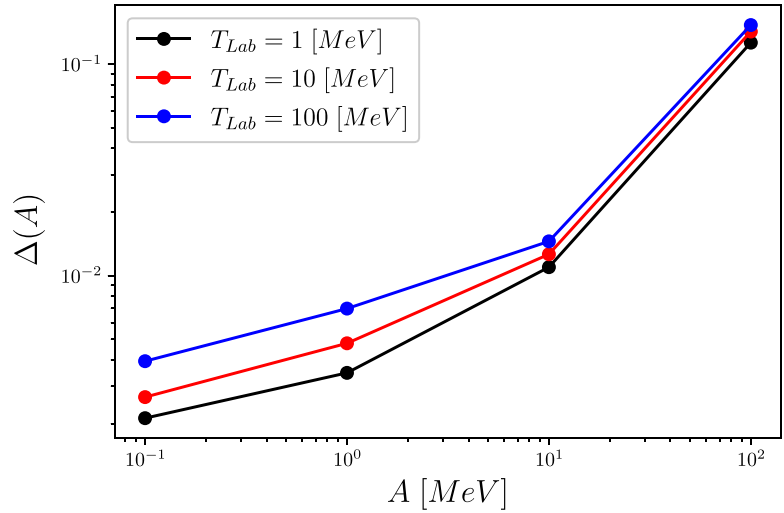

**Fig. 3.** Relative errors of the first-order Volterra approximation for bounded, vanishing potentials generated by random-phase multisines at $T_{\text{Lab}} = 1, 10, 100$ [MeV] laboratory energies.

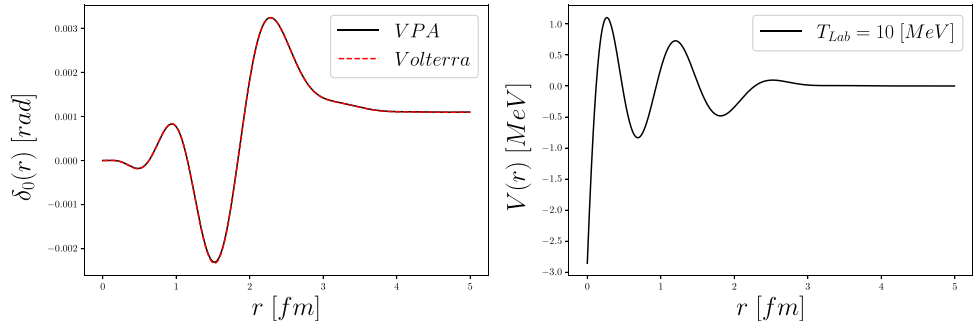

**Fig. 4.** Phase function of the first-order Volterra approximation compared to the phase function obtained from solving the VPA for a test potential shown in the right panel at $T_{\text{Lab}} = 10$ MeV.

laboratory energy is under a few hundred MeV. In Fig. 4 an example is shown for nucleon–nucleon scattering at $T_{\text{Lab}} = 10$ MeV, where the first-order approximation is good enough to describe the scattering system.

In nucleon–nucleon scattering, the potentials could be larger than a few tens of MeV [30–32], especially if we consider a hard repulsion near the scattering center. In that case, we would need to extend our approximation by including higher-order terms, or we would need to model the "remaining error" by some static nonlinear model if possible. To model nuclear potentials, we will use the latter method, but before that, in the next section, the inversion procedure will be shown, which will ultimately be used to describe the neutron–proton interaction potentials.

### 3.3. *Inversion of the first-order Volterra approximation*

In this section, we will use the first-order Volterra approximation to solve the inverse scattering problem, where the interaction potential is sought through the measured (or, in this case, simulated) phase shifts [33]. The forward problem is described by the VPA, and for the inversion, we will use the first-order Volterra representation given by a convolutional integral with the kernel function $h_1(z; r)$. In this section, we will use the previously determined operating range to be able to estimate the phase function using the BLA with good accuracy. Keeping all that in mind, we will approximate the s-wave phase function in the VPA equation by the first-order Volterra approximation as follows:

$$\delta_0(r, k) \simeq \int_0^r h_1(z; r, k) V(r - z) dz$$
$$= \int_0^r \frac{2\mu}{\hbar^2} \sum_{i=1}^N \sum_{j=0}^{2i} \left\{ \frac{(-1)^{i+j} 2^{2i-1}}{j!(2i-j)!} k^{2i-1} r^{2i-j} z^j \right\} V(r - z) dz, \quad (61)$$

where $h_1(z; r)$ is given by Eq. (55). In measurements, only the asymptotic $r \to \infty$ phase shifts are accessible, therefore, here we will set $r$ to a value where the potential is zero or very close to zero, and it will not cause any more detectable change in the phase function. The convolution integral in Eq. (61) could be inverted by Fourier transform and division [34]; however, when there is noise in the measurements, this simple technique is not very effective due to the sensitivity of the deconvolution operation [35]. As we would like to give a more robust and noise-insensitive approximation to the inversion problem, we will use a different method to estimate the $V(r)$ potentials from the measured asymptotic phase shifts. To solve the deconvolution problem, we will assume that the potentials we seek can be approximated by a finite order of continuous basis functions in coordinate space, in which case the integrals could be done analytically. In this section, we will use two different basis functions to solve the deconvolution: (1) nonorthogonal polynomials, and (2) orthogonal Legendre polynomials.

Let us start with the simplest representation, when the potential is given by a linear combination of nonorthogonal polynomials:

$$V(z) = \sum_{m=0}^M a_m z^m, \quad (62)$$

where $a_m$'s are the coefficients of the $m$'th basis functions, and the summation goes from 0 to $M$, where $M$ represents the highest order in the series. By substituting the nonorthogonal polynomials back into Eq. (61) we will arrive at the following form:

$$\delta_0(r, k) \simeq \int_0^r \frac{2\mu}{\hbar^2} \sum_{i=1}^N \sum_{j=0}^{2i} \left\{ \frac{(-1)^{i+j} 2^{2i-1}}{j!(2i-j)!} k^{2i-1} r^{2i-j} z^j \right\} \sum_{m=0}^M a_m (r - z)^m dz, \quad (63)$$

where the linear dependence on the unknown $a_m$ coefficients is apparent. The integral consisting of $z$ can be expressed analytically due to the simple polynomial form as:

$$\int_0^r z^j (r - z)^m dz = \frac{\Gamma(j+1)\Gamma(m+1)}{\Gamma(j+m+2)} r^{j+m+1}, \quad (64)$$

where, due to the fact that $j$ and $m$ are integer numbers, the gamma functions can be simply expressed as factorials. By putting back the integrated form into Eq. (63) we will arrive at the following analytical representation of the first-order Volterra response of the phase function when the potential is expressed with nonorthgonal polynomials:

$$\delta_{\text{polynomial}}(r, k) = \frac{2\mu}{\hbar^2} \sum_{m=0}^M a_m \sum_{i=1}^N \sum_{j=0}^{2i} \left\{ \frac{(-1)^{i+j} 2^{2i-1} m!}{(2i-j)!(j+m+1)!} k^{2i-1} r^{2i+m+1} \right\}, \quad (65)$$

where $\delta_{\text{polynomial}}(r, k)$ is the first-order Volterra approximation of $\delta_0(r, k)$ with nonorthogonal polynomial basis functions of order $M$. This form is clearly linear in the $a_m$ coefficients, which greatly simplifies the inversion procedure as the sought coefficients could be described in an analytical fashion or estimated by well-known optimization methods, e.g. gradient methods. We will describe the inversion procedure at the end of this section, but before we do that, let's express the first-order Volterra response with other basis functions, which will be more suited for our purpose as well.

It is well known that nonorthogonal polynomials are not the best choice for fitting purposes due to the large condition numbers of the coefficient matrices [36]. Large condition numbers are undesirable if one works with noisy data, therefore, we would like to constrain them as much as possible to be able to make a more robust estimation of the unknown coefficients. To do this, let's express the potential by Legendre polynomials [37] as:

$$V(z) = \sum_{m=0}^{M} a_m \mathcal{P}_m(z), \qquad (66)$$

where $\mathcal{P}_m(z)$ is the scaled $m$-th order Legendre polynomial. To be able to describe functions on $z \in [0, r]$ we need to scale the original $\mathcal{P}_m^{(0)}(z_0)$ Legendre polynomials defined as:

$$\mathcal{P}_m^{(0)}(z_0) = \frac{1}{2^m} \sum_{l=0}^{[\frac{m}{2}]} (-1)^l \binom{m}{l} \binom{2m-2l}{m} z_0^{m-2l}, \qquad (67)$$

where $z_0 \in [-1, 1]$ is the interval of the unscaled Legendre polynomials. To proceed, we need to apply a $z_0 \in [-1, 1] \to z \in [0, r]$ transformation, which can be done by $z_0 = \frac{2}{r}z - 1$. The scaled Legendre polynomials are now defined on $z \in [0, r]$ and can be expressed as:

$$\mathcal{P}_m(z) = \frac{1}{2^m} \sum_{l=0}^{[\frac{m}{2}]} (-1)^l \binom{m}{l} \binom{2m-2l}{m} \left(\frac{2}{r}z - 1\right)^{m-2l}. \qquad (68)$$

By substituting the scaled Legendre polynomials back into Eq. (61), we arrive at the following expression for the first-order Volterra response:

$$\delta_0(r, k) \simeq \int_0^r \left[ \frac{2\mu}{\hbar^2} \sum_{i=1}^{N} \sum_{j=0}^{2i} \left\{ \frac{(-1)^{i+j} 2^{2i-1}}{j!(2i-j)!} k^{2i-1} r^{2i-j} z^j \right\} \right.$$
$$\left. \times \sum_{m=0}^{M} \frac{a_m}{2^m} \sum_{l=0}^{[\frac{m}{2}]} (-1)^l \binom{m}{l} \binom{2m-2l}{m} \left(\frac{2(r-z)}{r} - 1\right)^{m-2l} \right] dz, \qquad (69)$$

where the integrals are now a bit more complicated than in the nonorthogonal case; however, they can be put into a more convenient form by applying the solution of the following integral:

$$\int_0^r z^j \left(\frac{2(r-z)}{r} - 1\right)^{m-2l} dz = {}_2F_1\left[j+1, 2l-m; j+2; 2\right] \frac{r^{j+1}}{j+1}, \qquad (70)$$

where ${}_2F_1[a, b; c; d]$ is the hypergeometric series [38], and again the integral has the form $\sim r^\alpha$, where $\alpha$ is a positive integer number, which depends on the number of terms in the Taylor expansion of the $\sin^2()$ nonlinearity in the original differential equation. Putting everything back into Eq. (69), the first-order Volterra response, using scaled Legendre polynomials can be expressed as:

$$\delta_{\text{Legendre}}(r, k) = \frac{2\mu}{\hbar^2} \sum_{m=0}^{M} a_m \sum_{i=1}^{N} \sum_{j=0}^{2i} \sum_{l=0}^{[\frac{m}{2}]} \left\{ \frac{(-1)^{i+j+l} 2^{2i-m-1}}{j!(2i-j)!(j+1)} \binom{m}{l} \binom{2m-2l}{m} \right.$$
$$\left. \times {}_2F_1\left[j+1, 2l-m; j+2; 2\right] k^{2i-1} r^{2i+1} \right\}, \qquad (71)$$

where, despite the fairly complicated expression with the hypergeometric series, the actual result for relatively low-order polynomials is quite straightforward and easy to obtain. The first-order response $\delta_{\text{Legendre}}(r, k)$ will have much better behavior than the previously derived $\delta_{\text{polynomial}}(r, k)$, which will be apparent from the obtained condition numbers at the end of this section.

After showing that the convolutional integral in the first-order Volterra response can be expressed in closed form by using polynomial basis functions, let's proceed with the actual inverse problem we would like to solve. Here, let's assume that the scattering problem is in the operating range of the first-order Volterra model, which means that the sought potential should be in the range of a few tens of MeV's and it should vanish around $r \approx 5$ fm. These numbers are the consequence of the convergence studies shown in the previous subsection. As we have seen before, each case has a linear dependence on the identifiable $a_m$ coefficients, therefore, the problem is greatly simplified in contrast to the case when the system is described by a higher-order Volterra series.

In the general case, when the phase function at some specific energy ($k_j$) is described by a higher-order Volterra model and the potential is sought in the form of a linear combination of some basis functions, e.g. orthogonal polynomials, the response can be cast in the following simplified form:

$$\delta_0(r, k) \simeq \sum_{m=0}^{M} a_m \text{H}_1(r, k; N, m) + \sum_{m_1=0}^{M} \sum_{m_2=m_1}^{M} a_{m_1} a_{m_2} \text{H}_2(r, k; N, m_1, m_2)$$

$$+ \sum_{m_1=0}^{M} \sum_{m_2=m_1}^{M} \sum_{m_3=m_2}^{M} a_{m_1} a_{m_2} a_{m_3} \text{H}_3(r, k; N, m_1, m_2, m_3) + \cdots, \quad (72)$$

where the $k$ momentum variable naturally corresponds to the energy of the scattering process, while the $\text{H}_n(r, k; N, m_1, \ldots m_n)$ functions contain the $n$-dimensional convolutional integrals, which depend on the parameter $N$ (the order of the Taylor expansion of the $\sin^2()$ nonlinearity). If one keeps only the first term, we arrive at the previously described BLA of the nonlinear dynamical system, which in this case means the forward problem described by the VPA equation. As our task is to estimate the interaction potential by measuring the asymptotic phase shifts, we need to find the inverse of the functional form shown in Eq. (72), which in this case boils down to finding the unknown $a_i$ expansion parameters by knowing the $\delta_0(\hat{r}, k_j)$ phase shifts at some discrete energies corresponding to $k_j$ momenta, and at a "far enough" $\hat{r}$ coordinate, which is related to the asymptotic phase shift where after that point the phase function will not have significant changes. As the Volterra representation has a simple polynomial form, the overall inverse problem is now a well-known optimization problem that can be solved by multiple numerical methods, even for higher-order polynomials that correspond to higher-order Volterra representations.

Here, we will try to solve the inverse problem by using only the first-order Volterra representation, which means the problem can be cast into the following matrix form:

$$\begin{bmatrix} \text{H}_1(\hat{r}, k_1; N, 1) & \text{H}_1(\hat{r}, k_1; N, 2) & \cdots & \text{H}_1(\hat{r}, k_1; N, M) \\ \text{H}_1(\hat{r}, k_2; N, 1) & \text{H}_1(\hat{r}, k_2; N, 2) & \cdots & \text{H}_1(\hat{r}, k_2; N, M) \\ & & \vdots & \\ \text{H}_1(\hat{r}, k_L; N, 1) & \text{H}_1(\hat{r}, k_L; N, 2) & \cdots & \text{H}_1(\hat{r}, k_L; N, M) \end{bmatrix} \begin{bmatrix} a_1 \\ a_2 \\ \vdots \\ a_M \end{bmatrix} = \begin{bmatrix} \delta_0(\hat{r}, k_1) \\ \delta_0(\hat{r}, k_2) \\ \vdots \\ \delta_0(\hat{r}, k_L) \end{bmatrix}, \quad (73)$$

where $L$ represents the number of asymptotic phase shifts we can use for the inversion, $k_1, k_2, \ldots, k_L$ are the corresponding energies, while $\text{H}(\hat{r}, k; N, m)$ are the first-order convolutional integrals with the corresponding parameters at a specific $\hat{r}$ coordinate. The linear system of equations representing the forward system is now directly invertible by matching the number of measured phase shifts ($L$) and the order of the expansion of the potential ($M$), or by using all the available phase shifts (which could be more or less than the order of the expansion in the potential), the $a_m$ parameters could be estimated by numerical techniques, e.g. by the conjugated gradient method. A good measure for the robustness of the inversion in this case would be the condition number of the coefficient matrix, which will depend on the type of expansion of the V($r$) potential function. One would expect that the condition number of the orthogonal polynomial expansion will be smaller than in the nonorthogonal case, which indeed will be the case; however, it is not straightforward to see due to the rather complicated expressions appearing in the Taylor expansion of the nonlinearity. Due to our a priori knowledge of the physical system, it is easy to set up well-defined constraints for the optimization problem. To show the general behavior of the method that we will extend later on to describe real-life physical systems, let us set up a benchmark problem first, where we will try to estimate the following potential from the measured phase shifts:

$$V(r) = e^{-2r}(r^4 - 1), \tag{74}$$

where the exponential decaying factor makes sure that the potential is within the operating range of the first-order Volterra approximation. To construct the coefficient matrix, let us calculate the phase shifts in the energy range $T_{\text{Lab}} \in [1, 101]$ MeV with a resolution of $\Delta T_{\text{Lab}} = 5$ MeV, which means $L = 21$ discrete points, which can be used to determine the unknown $a_m$ coefficients. The momentum $k$, which appears in the derivations, can be determined from the laboratory energy as:

$$k = \frac{m_2^2 \left(T_{\text{Lab}}^2 + 2m_1 T_{\text{Lab}}\right)}{(m_1 + m_2)^2 + 2m_2 T_{\text{Lab}}}, \tag{75}$$

where $m_1$ and $m_2$ are the masses of the scattering particles set to $m_1 = m_2 = 940$ MeV. The next step in the inversion is the choice of the optimal order $M$ of the polynomials. This is not a straightforward task, as a too-complex model tends to overfit, and the coefficient matrices could have large condition numbers, which will be problematic if the measured phase shifts have too much noise. In contrast, a too-simple model will not be able to grasp the fine structure of the potential, and we will not be able to reproduce the measured phase shifts well enough. This is, of course, a well-known problem, and in practice, a few methods exist that use different criteria to assess this issue, e.g. Akaike information criterion (AIC) and Bayesian information criterion (BIC) [39,40].

Here, we will use a more straightforward method to deduce the optimal order of the polynomial approximations, which will take into account that the original system is described by not the Volterra model, but the VPA. In this way, we are able to compare the obtained phase shifts calculated by the VPA equation from the estimated (inverted with the Volterra model) potentials to the original phase shifts. By doing that, we will implicitly add some extra information to the order selection procedure, e.g. if the inverted potential is outside of the assumed operating range of the first-order Volterra approximation, then this will be shown on the recalculated phase shifts because, in this case, the potential we obtain from inversion is not guaranteed to be part of the forward system that is estimated by the first-order Volterra approximation. Therefore, to determine the optimal order of the polynomial approximations, we will compare the original and the recalculated phase shifts by calculating their relative errors at different orders

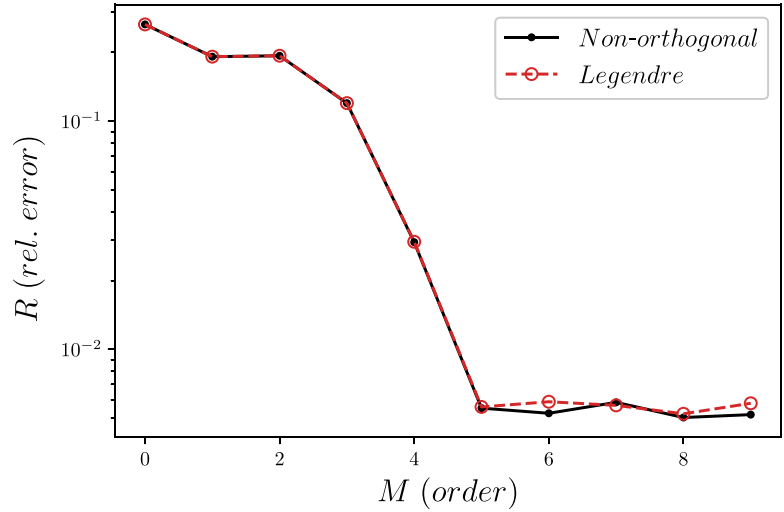

**Fig. 5.** Relative errors of the recalculated phase shifts from the inverted potentials with non orthogonal and Legendre polynomial expansions of different orders. In both cases the optimal order is $M = 5$.

as follows:

$$R = \frac{1}{L}\sum_{i=1}^{L} \frac{|\phi^{\text{true}}(k_i) - \phi^{\text{est}}(k_i)|}{|\phi^{\text{true}}(k_i)|}, \qquad (76)$$

where $L$ is the number of phase shifts at different $k_i$ momenta, while $\phi^{\text{true}}(k_i)$ is the true asymptotic phase shift, and $\phi^{\text{est}}(k_i)$ is the estimated phase shift obtained by plugging the inverted potential function into the VPA equation and determining the asymptotic phase shifts at all the necessary $k_i$ momenta.

Using the above definition, after constructing the coefficient matrices in Eq. (73), inverting them, and using the obtained potentials in the VPA equation at the previously defined $L = 21$ energies, we get the results shown in Fig. 5 for the nonorthogonal, and for the Legendre polynomial cases, respectively. The relative error calculated in this way shows the same behavior for both cases, which is not surprising as both are polynomial approximations. The huge difference will be in their respective condition numbers of the coefficient matrices, which can be seen in Fig. 6. The condition numbers shown in Fig. 6 suggest that our initial guess was indeed correct, as the Legendre polynomial expansion has much lower values at all orders than its nonorthogonal counterpart. This will have great consequences when dealing with noisy data, but before that, let's show our results with the derived optimal orders $M = 5$ for each case. In Figs. 7 and 8 the estimated potentials with the recalculated phase shifts can be seen for the nonorthogonal, and the Legendre polynomial cases, respectively. As can be seen, both potentials are very similar, which is not a surprise as they both have similar polynomial approximations. In both figures, the right panel shows the original and recalculated phase shifts. In that graph, the blue circles (Volterra) show the phase shifts calculated by the first-order Volterra approximation with the inverted potentials, while the red crosses (Inversion (VPA)) show the phase shifts calculated by the VPA equation with the inverted potentials. In this case, both results are very close to each other, which is the expected behavior when the inverted potential is in the operating range of the Volterra approximation.

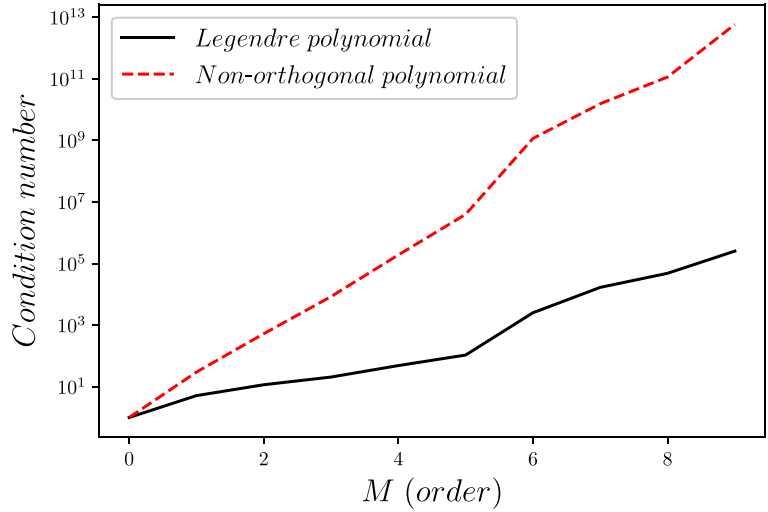

**Fig. 6.** Condition numbers of the coefficient matrices defined in Eq. (73) for the nonorthogonal and Legendre polynomial cases. The graph strongly suggests that the Legendre polynomial expansion will be much better in describing noisy data than will the nonorthogonal case.

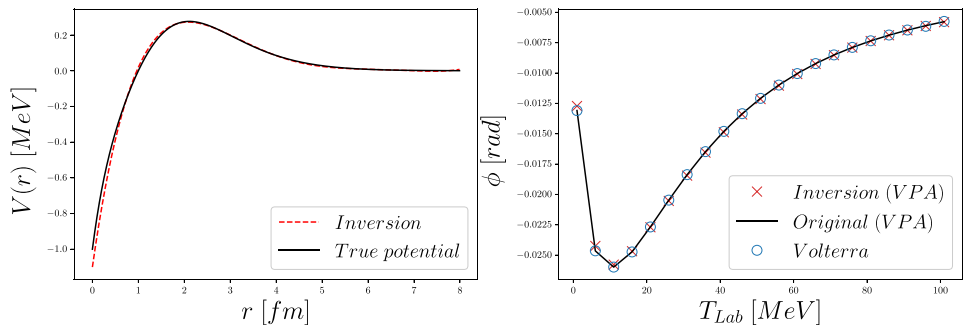

**Fig. 7.** Reconstructed potential (left) and the corresponding phase shifts (right) using nonorthogonal polynomials as basis functions. The phase shifts marked by circles correspond to the values obtained by using the first-order Volterra approximation, while the red crosses show the values for the VPA calculations with the same inversion potential.

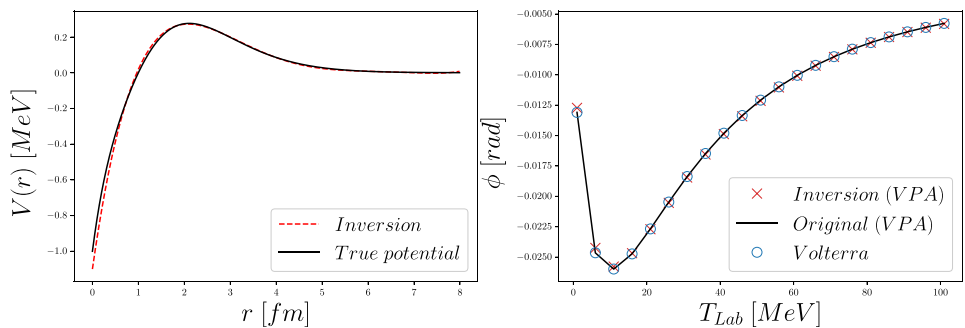

**Fig. 8.** Reconstructed potential (left) and the corresponding phase shifts (right) using Legendre polynomials as basis functions.

To conclude this section, we will check the robustness of the inversion with the Legendre expansion by introducing some reasonable noise into the phase shifts used in the inversion procedure. To do this, let's add uniformly distributed random noise to each phase shift and do the inversion with the noisy data. Figure 9 shows the potentials obtained with the recalculated and original phase shifts when the random noise corresponds to 40% (upper side) and 20% (bottom side) relative errors measured relative to the original phase shift data. The results are convincing in both cases, as the general characteristics of the potentials are reconstructed even when the noise is relatively large. From the recalculated phase shifts, it can be seen that the Volterra approximation tends to smooth out the general behavior of the energy dependence of the phase shifts, which will manifest itself in the obtained smooth potential functions. In the next section, we will extend the model to be able to describe potentials that are outside of the operating range of the first-order Volterra approximation and use the final model to obtain the nuclear potentials using the measured nucleon-nucleon scattering data.

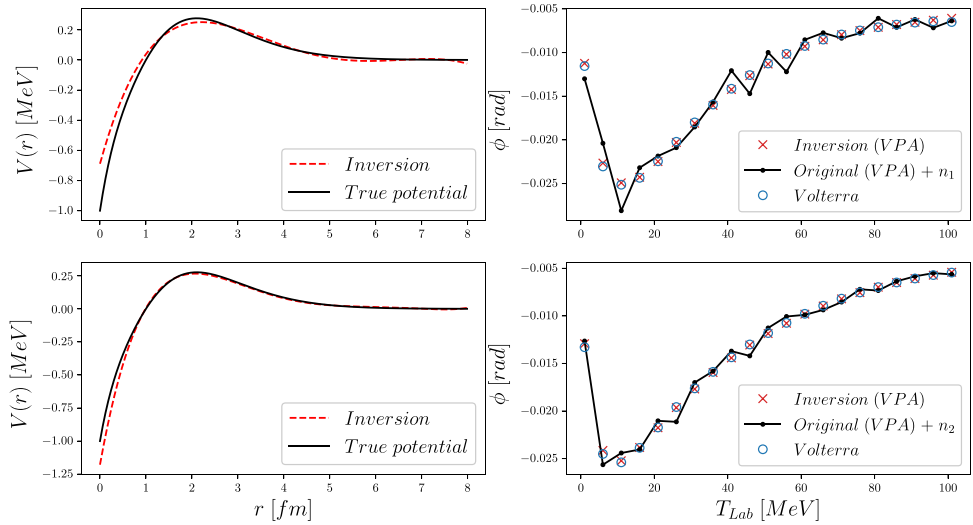

**Fig. 9.** Reconstructed potential (left) and the corresponding phase shifts (right) using Legendre polynomials as basis functions, obtained by using noisy phase shift inputs.

## 4. Nuclear potentials from the first-order Volterra expansion

In this section, the previously shown inversion technique with the first-order Volterra approximation will be applied to the nucleon-nucleon inverse scattering problem. To do so, however, the method has to be extended to be able to describe potentials which are expected to be out of the operating range of the first-order approximation. One way to do this would be to use higher-order terms in the Volterra approximation; however, in this case, the inversion procedure will lose its simplicity because we will not have a simple linear relationship between the free parameters and the measured phase shifts. It is worth mentioning, however, that when the system is described by higher-order Volterra approximations, instead of Eq. (73) we will arrive at a polynomial relation between the free parameters and the phase shifts, which can be solved numerically by many well-known methods [41].

Here, however, we will use a different approach, mainly due to the simplicity and robustness of the first-order approximation, which have been shown in the previous section. Let us first make a few assumptions about the system we would like to describe. On one side, the sought potentials fall into the previously described picture as they have to be (1) continuous and (2) vanishing near 4–5 [fm]. By examining the first-order Volterra approximation in the previous section, we have seen that the relative errors do not depend too much on the energy of the collisions; however, they have an increasing tendency with larger potentials. In Fig. 3 it can be seen that for potentials reaching a few hundred MeV's, the relative error is about 10%, which is not too large, but it could be problematic in practical applications. Our task is therefore to somehow model this difference between the first-order Volterra approximation and the true system in the operating range of potentials a few hundred MeV in magnitude and at collision energies below $T_{\text{Lab}} = 200$ MeV. While doing this, we would like to keep the first-order, linear relationship between the $a_m$ expansion parameters and the measured phase shifts, therefore, we will seek a (possibly) nonlinear transformation that transforms the phase shifts of the true system described by the VPA equation to the phase shifts given by the first-order Volterra approximation for the same potential function. To make this more clear, we aim to find the following transformation:

$$\phi_V(k) \simeq Q\Big(\phi_0(k), \dots \Big), \tag{77}$$

where $\phi_0(k)$ are the true phase shifts obtained by solving the VPA equation for some potential $V(r)$, while $\phi_V(k)$ is the first-order Volterra response for the potential and is given by the first term on the right-hand side in Eq. (72). By defining the problem this way, the system could be approximated by the following expression:

$$Q\Big(\phi_0(k), \dots \Big) \simeq \sum_{m=0}^{M} a_m H_1(\hat{r}, k; N, m), \tag{78}$$

where $H_1()$ consists of the convolutional integral, whose form depends on the type of expansion that we use for the potential functions e.g. nonorthogonal polynomial, Legendre polynomial, etc., while $\hat{r}$ refers to the cutoff distance. In general, the Q() function could be anything, and it is not necessary for it to be invertible. This way, we have kept the linear relationship between the $a_m$ parameters and the (now) transformed $Q(\phi_0(k), \dots)$ phase shifts.

To find the Q() transformation, we will use a data-driven approach and fit an RBF-type neural network in a desired operating range set by the applied training, validation, and test data, for which we generate 10 000 vanishing potentials by random-phase multisines defined in Eq. (60) in the magnitude range between $-200$ MeV and 200 MeV. The data have been divided as follows: 8000 samples are used for training, and 2000 samples are used for testing and validation purposes. It is also a modeling step on how to define the inputs and outputs of the sought transformation, e.g. for inputs, it could be simply the $\phi_0(k_i)$ phase shift at a specific $k_i$, or it could be the full $(\phi(k_1), \dots, \phi(k_n))$ set of phase shifts, while the output could also be only one or more phase shifts. By trying out different configurations, we came to the conclusion that to be able to give reasonably good estimations, it is necessary to set up a Multi Input Single Output (MISO) system, where the inputs of the system will be the $\phi_0(k_1), \phi_0(k_2), \dots, \phi_0(k_n)$ measured phase shifts at a predefined energy range and a specific $k_j$ parameter, while the output of the system will be a single $\phi_V(k_j)$ phase shift at the same $k_j$ energy. Considering all the above, the transformation we seek can be described by the following expression:

$$\phi_V(k_j) = \underbrace{\sum_{m=1}^{M_{\text{RBF}}} w_m e^{-\left[\sum_{i=1}^{n} \frac{(\phi_0(k_i)-b_i)^2}{\sigma_i} + \frac{(k_j-b_{n+1})^2}{\sigma_{n+1}}\right]}}_{Q(\phi_0(k_1),\phi_0(k_2),\dots,\phi_0(k_n),k_j)}, \tag{79}$$

where $w_m$, $b_i$, and $\sigma_i$ are the free parameters of the RBF network, which have to be trained through an appropriate method. As there is no need for this transformation to be able to describe the whole system at any operating range, the determination of the $k_1, \ldots, k_n$ grid could be problem-specific; however, it has to be noted that if one wants to describe a different system, e.g. in a different energy range, the network has to be retrained with the appropriate new parameters. In our case, the scattering problem is defined in the range approximately between $T_{\text{Lab}} = 1$ MeV and $T_{\text{Lab}} = 200$ MeV, which corresponds roughly to $k_i \in [0.1, 1.5]$ fm$^{-1}$. By trying out different configurations, it has been concluded that a uniform grid in $k_i$ is much better than a uniform grid in $T_{\text{Lab}}$, therefore, we set the $k_i$ parameter range to $k_i \in [0.1, 1.5]$ fm$^{-1}$ with a resolution of $\Delta k_i = 0.1$ fm$^{-1}$, which means $15 + 1$ inputs for the RBF network (15 phase shifts and 1 $k_j$). To train the free parameters $w_m$, $b_i$, and $\sigma_i$, the method described in Ref. [42] is used, where the $b_i$ centers and $\sigma_i$ widths are chosen in an unsupervised manner, and the $w_m$ parameters are determined in a supervised manner by solving the remaining linear system of equations.

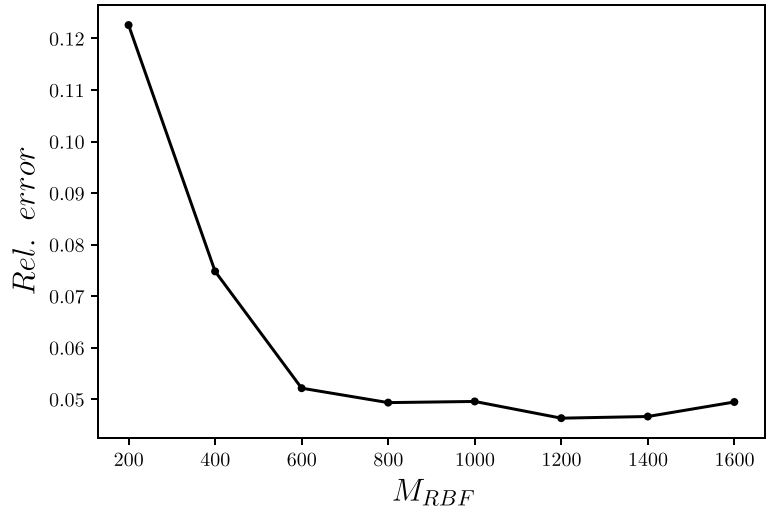

**Fig. 10.** Relative errors of RBF networks with different numbers of basis functions. The networks are trained by using 8000 training data, while the errors are calculated by using 2000 test data.

The complexity of the network ($M_{\text{RBF}}$) has been determined by trying out different configurations with different numbers of neurons in the hidden layer, whose results can be seen in Fig. 10, where the averaged relative errors for the test data can be followed as a function of the number of basis functions. By examining the errors, it can be concluded that $M_{\text{RBF}} \simeq 1000$ basis functions is enough to describe the system with an accuracy of a few percent ( 5%), which will be enough for the application in which we intend to use the model. From the calculated relative errors, we can also see the general tendency that the error will decrease when we add more neurons; however, after some point, the network will start to overfit (overtrain), and the test errors, which in this case were used for validation as well, will stagnate or even increase. By using more training data and more neurons, it would be possible to achieve a better generalization; however, for our applications, this accuracy to a few percent will suffice.

In Fig. 11 we show results of an RBF network with $M_{\text{RBF}} = 1000$ basis functions for 40 normalized test data, which means that the outputs (phase shifts) are normalized to the interval $[-1, 1]$. The training of the network has been done by first normalizing the inputs and outputs between $[-1, 1]$, then "training" the widths and centers in an unsupervised manner and finally solving the remaining linear system of equations to obtain the $w_m$ coefficients.

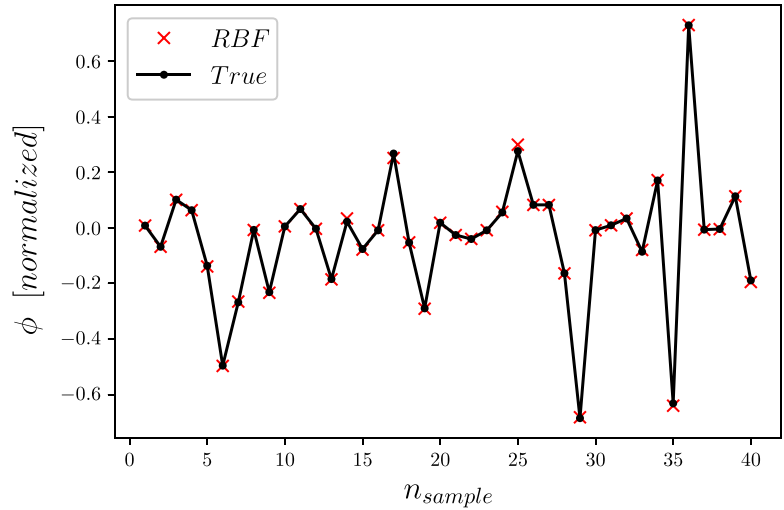

**Fig. 11.** Comparison of the output of the trained RBF network with $M_{\text{RBF}} = 1000$ basis functions (red crosses), to the test data (black line with dots). By considering Fig. 10 the relative errors are approximately 5%.

As can be seen, the RBF network was able to describe relatively well the remaining nonlinear errors after identifying and using the first-order Volterra approximation; however, some remarks are in order. Firstly, we have used vanishing potentials generated by multisine signals for training the RBF network. In general, we could have used any other type of training data that suit our needs, e.g. they do not have to be vanishing, or they could be rapidly oscillating, etc. It has to be noted, however, that with more complex signals, the resulting RBF network should be more complex as well. It is also possible that a simple "static" RBF construction is not enough to be able to describe the remaining dynamic nonlinear system.

On the other hand, it is not necessary to model every possible phase shift generated by the vanishing multisine potentials. In this case, we did not make any constraints on the possible output phase shifts that we wanted to model; however, by using potentials that correspond to a specific interval of phase shifts, e.g. $\phi \in [0, \pi]$, the network could be greatly simplified. This could be simply done by knowing the phase shifts we want to describe.

By applying the trained RBF network, in theory, now we would be able to model a nucleon–nucleon scattering scenario with a few percent accuracy if the potential is approximately between $[-200, 200]$ MeV. This assumption is, however, not neccessarily known before we make some tests or know more about the system in question. One thing we could do is check the generated phase shifts of the training data, and if they cover the range we would like to describe, then there is a huge possibility that a potential which satisfies the constraints can be found.

To be able to give a more precise estimation for the interaction potential, we will introduce a correction step into our inversion scheme so that we will be able to make a better fit to the phase shifts by making small changes in the potentials. First, let's assume that after the inversion, we have obtained a potential $V_0(r)$, which is able to describe the measured phase shifts with a few percent accuracy. By assuming that the general shape of the potential is readily given by this first estimation, we only want to make small corrections in this potential as $V_{\text{corr}}(r) = V_0(r) + \Delta V(r)$. The correction will be given by using Spline basis functions by first interpolating the starting potential $V_0(r)$ with a predetermined number of control points. The number and places of control points can be easily determined by observing the $V_0(r)$ potentials, e.g. if the potential should be described in the range of $r \in [0, 6]$ fm, then a good choice would be 10–20 control points distributed uniformly between 0 fm and 6 fm.

Next, to make the correction step, we will vary the "amplitude" of the control points until we reach an optimum or a satisfying value in the following error function:

$$E = \frac{1}{N_\phi} \sum_{i=1}^{N_\phi} \frac{|\phi(k_i^{\text{meas}}) - \phi_{\text{corr}}(k_i^{\text{meas}})|}{|\phi(k_i^{\text{meas}})|}, \tag{80}$$

where $N_\phi$ is the number of phase shifts we use for the correction step at the original (measured) and not the interpolated momentum grid $k_i^{\text{meas}}$, $\phi(k_i^{\text{meas}})$ are the measured phase shifts, while $\phi_{\text{corr}}(k_i^{\text{meas}})$ are the recalculated phase shifts from the corrected $V_{\text{corr}}$ potential. To optimize for this error function, we will use the Simulated Annealing (SA) algorithm, which has some probability of accepting corrections with larger errors; however, this acceptance probability will decrease in time with a predetermined speed given by the following exponential form:

$$T = T_0 e^{-kn}, \tag{81}$$

where $T_0$ and $k$ are free parameters, while $n$ is the number of steps in the annealing algorithm. The free parameters control the speed of convergence and are problem-dependent. By applying the previously described method, the full inversion scheme can be summarized in the following three steps:

- Interpolate the measured phase shifts $\phi_0$ to the predetermined momentum grid (used for the RBF network) and transform the interpolated phase shifts into the Volterra phase shifts $\phi_V$ by using the trained RBF network using Eq. (79).
- Choose suitable basis functions, and an order $M$ for the potential (polynomial, Legendre, etc.), and build the coefficient matrix in Eq. (73), then solve the resulting linear system of equations for the unknown $a_m$ parameters described by Eq. (78).
- Assuming the obtained potential $V_0(r)$ is close to the real one, make small corrections on it by expanding the $V_0(r)$ potential with Spline basis functions, and using the SA algorithm to try to find a better solution $V_{\text{corr}}(r)$, which has a better fit to the measured phase shifts at the orginal (not interpolated) momentum grid.

Finally, we could use the described method to obtain the interaction potentials in nucleon–nucleon scattering at fixed angular momentum, for which we will use the data for the $^1S_0$ neutron–proton phase shifts summarized in Ref. [43]. After going through all the steps of the inversion scheme (interpolating the measured phase shifts onto the predetermined momentum grid, solving the linear system of equations for the coefficients of the polynomials, then making small corrections by the corrector step), we arrive at the interaction potential shown in Fig. 12. The obtained potential has the expected shape, with a stronger repulsive part at short distances, possibly due to nonperturbative quantum chromodynamic effects, and a long-distance attractive part, while the whole potential tends to vanish around 5 fm.

Next, let us check the recalculated phase shifts for the uncorrected and also for the corrected potentials. This can be followed in Fig. 13, where in the left panel the uncorrected phase shifts calculated at the interpolated momentum grid are compared to the measured and interpolated phase shifts, while in the right panel the corrected phase shifts now at the measured momenta are compared to the measured phase shifts. In the left panel, the comparison has been done on the interpolated momentum grid because the pure inversion is done at those $k_i$ values; however, the correction step uses the original momentum grid, therefore, in the right panel, the corrected results are compared at those momenta. From the comparison of the phase shifts, it can be seen that the pure inversion was able to reproduce the measured phase shifts relatively well with a few percent averaged accuracy, which is the expected behavior, since the RBF network, which models the difference between the first-order Volterra approximation and the full nonlinear dynamical system, had an accuracy of the same order of magnitude. After the correction, the relative error became even smaller, reaching less than 1%, which can be followed in Fig. 14, where the averaged relative errors are shown at each accepted annealing step. Finally, let's check the difference between the pure and corrected potentials. This can be seen in Fig. 15, which shows that indeed, a few MeV corrections here and there will give us the necessary corrections in the phase shifts. This means that the crude shape of the sought potentials is indeed achievable without using any correction step, and after we have a good first guess of the potentials, any other suitable method could be used to achieve an even better fit to the measured phase shifts.

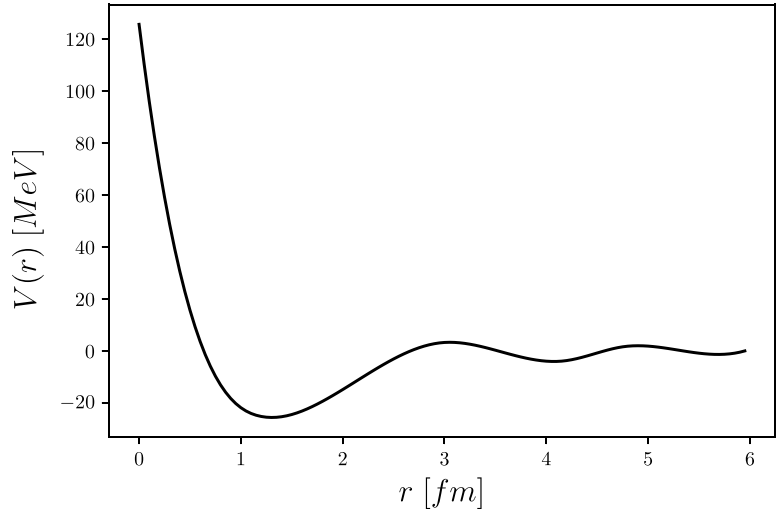

**Fig. 12.** The final $V_{\text{corr}}(r)$ potential obtained after inversion using 5th-order nonorthogonal polynomials.

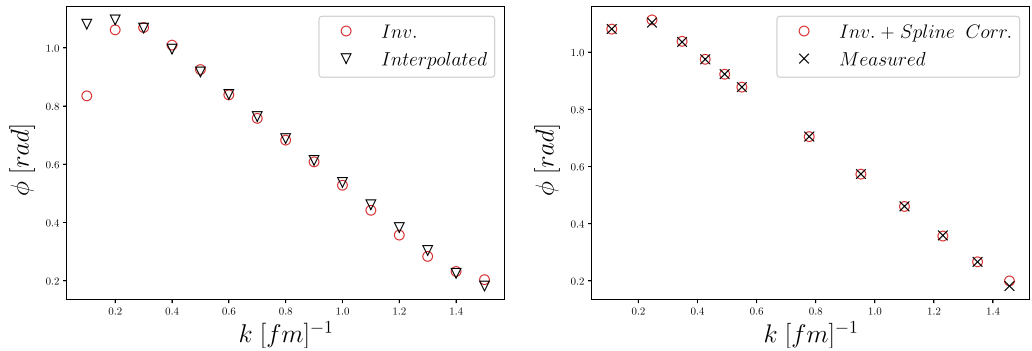

**Fig. 13.** Comparison of the recalculated phase shifts using the obtained pure $V_0(r)$ and corrected $V_{\text{corr}}(r)$ potentials. In the left panel the comparison is done between the uncorrected results and the measured phase shifts interpolated to the momentum grid of $k_i \in [0.1, 1.5]$ with a resolution of $\Delta k_i = 0.1$, while in the right panel the corrected results are shown compared to the measured phase shifts at $k_i^{\text{meas}}$.

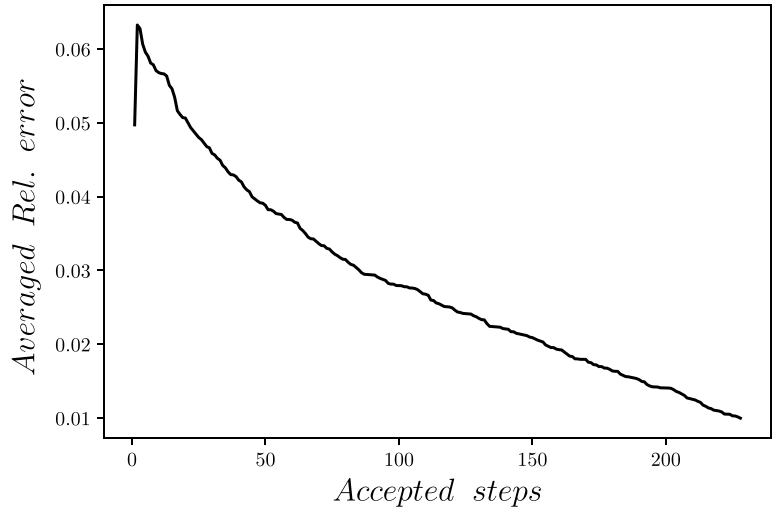

**Fig. 14.** Averaged relative errors during the SA optimization with a stopping condition of 1% relative error. The initial 5% value corresponds to the recalculated phase shifts using the uncorrected $V_0(r)$ potential and calculated at the $k_i^{\text{meas}}$ momenta.

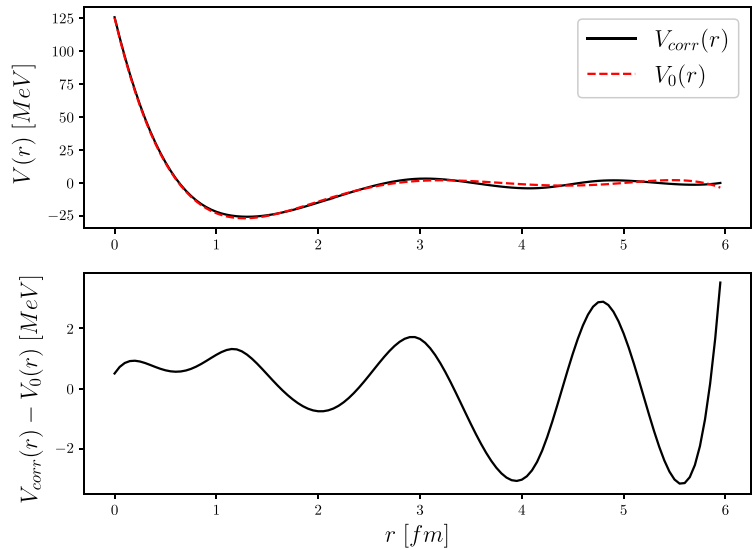

**Fig. 15.** Comparison of the pure $V_0(r)$ and the corrected $V_{\text{corr}}(r)$ potentials. The upper panel shows the potentials, the lower panel shows their difference.

With this simple real-life example, we have shown that the method is indeed able to describe the inverse nonlinear and nonautonomous system with good accuracy, even without final state correction. The main advantage of the method is the analytical representation of the underlying dynamical system for weak potentials, which can be extended by additional terms in the Volterra series. Here, we only applied the first-order Volterra representation, which gave us the BLA of the nonlinear system, therefore, the inverse problem boils down to solving a 1D deconvolutional problem. This approximation is only valid at a certain operating range; therefore, if we want to use the Volterra method to solve the more general inverse problem, we have to add higher-order terms into the Volterra series or have to model the difference between the full system and the BLA with some additional method. In the first case, the Volterra kernels can be determined in an analytical fashion, and after we choose a basis function set for the expansion of the sought potentials (e.g. polynomials), the inverse problem can be formulated as an algebraic (polynomial) system of equations that has multiple well-developed numerical methods for its solution. In the second method, the difference between the true and the first-order system has to be modeled by a suitable method, e.g. by using neural networks. Here, an RBF-type neural network has been used to model the "remaining" system. The advantage of this method is the flexibility in how to model the difference, as we can easily choose the inputs, outputs, and hidden parameters, and we can even use the preliminary knowledge of the system to constrain the training samples, therefore greatly simplifying the possible model.

The method described before should be able to model the system with an accuracy of approximately the applied neural network model, therefore, it is a crucial step to fit a model that is able to generalize well enough to inputs that were not part of the training samples. Solving the inverse problem this way, we will be able to obtain a good representation of the sought potential; however, as the phase shifts are very sensitive to even small perturbations in the potentials, whenever we might have some larger discrepancy in some of the phase shifts, we could apply a correction step to "fine-tune" the potentials. In the case of the $^1S_0$ neutron-proton scattering example, we have seen some larger discrepancies between the phase shifts only at lower energies; therefore, we could have assumed that the $V_0(r)$ potential needed some minor corrections at larger distances near its decaying tail, which was indeed the case.

## 5. Conclusions

In this paper, the Volterra series method to approximate nonlinear dynamical systems has been extended to be able to describe nonautonomous nonlinear differential equations as well. In this case, to be able to determine the Volterra kernels, differential equations have to be solved instead of the usual algebraic equations. The method has been used to obtain the first-order Volterra kernel of the VPA with zero angular momentum, which is a first-order nonautonomous and nonlinear differential equation used to describe the phase shift evolution in an s-wave two-body elastic quantum scattering scenario. We have shown that the first-order Volterra approximation is able to describe the system in a reasonably wide operating range; however, to be able to describe nuclear scattering, it was necessary to include higher-order Volterra terms or model the difference between the full system and the linear approximation with an appropriate model. We chose to describe the difference as a nonlinear noise term with RBF neural networks. By using reasonable assumptions on the potentials (vanishing, continuous, bounded, etc.), we were able to model the noise term as a "static" system in the sense that it does not depend on the coordinates $r$ but only on multiple phase shifts at a predetermined momentum grid. By extrapolating the measured phase shifts to this momentum grid and then using the RBF network to transform them into some new phase shifts, which can now be described by the first-order Volterra approximation, the inverse problem can be cast into a 1D deconvolutional form. By expanding the sought-after potentials in polynomials, the deconvolution can be written as a linear system of equations for the unknown coefficients, which can be easily solved. To determine the optimal order of the polynomials, we recalculated the phase shifts by solving the VPA equation for the resulting potentials at each order. After the inversion, the potential should be able to approximate the true one with the approximate accuracy of the applied neural network. At the last stage, a fine-tuning step has been applied using Spline basis function interpolation and SA optimization, whereby, using small adjustments in the values at the control points, we were able to make small, continuous changes in the potentials in coordinate space. This method can be used for optimization by using the averaged relative difference of the measured and recalculated phase shifts as a goal function for the SA method.

The described method has been applied to the $^1S_0$ neutron-proton scattering at fixed angular momentum in the energy range of $1 < T_{\text{Lab}} < 200$ MeV, giving very good results with an averaged relative error of less than 1% at the measured phase shifts. Regarding the inverse scattering problem the method could be extended to be able to describe nuclear scattering at fixed energy, where the general, angular momentum–dependent VPA equation has to be approximated by the nonautonomous Volterra series method. It is also possible to include singular potentials into the picture; however, to do so, we have to consider different initial conditions when determining the Volterra kernels, which could complicate the analytical determination of the higher-order kernels.

**Funding**

This work was supported by the National Research Foundation of Korea (NRF) grant funded by the Korea government (MSIT) (No. 2018R1A5A1025563). The author was supported by the Hungarian OTKA fund K138277.